\documentclass{jfm}
\usepackage{color}
\usepackage{graphicx}
\usepackage{natbib}
\usepackage{amsmath}
\usepackage{caption}
\usepackage{subcaption}

\usepackage{booktabs}
\usepackage[table,xcdraw]{xcolor}
\usepackage{multirow}
\usepackage{rotating,tabularx}

\newcommand{\NOTE}[1]{{\color{black}{#1}}}
\newcommand{\NOTEN}[1]{{\color{black}{#1}}}

\title{\NOTE{Buoyancy driven bubbly flows: scaling of velocities in bubble columns operated in the heterogeneous regime}}

\author[Y. Mezui, M. Obligado and A. Cartellier]%
       {Y. Mezui\aff{1}, M. Obligado\aff{1} and A. Cartellier\aff{1}
         \corresp{Email address for correspondence:  alain.cartellier@cnrs.fr}}

\affiliation{\aff{1} Universit\'{e} Grenoble Alpes, CNRS, Grenoble-INP, LEGI, F-38000, Grenoble, France }

\pubyear{2010}
\volume{650}
\pagerange{119--126}
\date{?; revised ?; accepted ?. - To be entered by editorial office}

\begin{document}
\sloppy 

\maketitle
\begin{abstract}

The hydrodynamics of bubble columns in the heterogeneous regime is revisited. Focusing on air-water systems at large aspect ratio, we show from dimensional analysis that buoyancy equilibrates inertia, and that velocities scale as $(gD\varepsilon)^{1/2}$, where $D$ is the bubble column diameter, $\varepsilon$ the void fraction and $g$ the gravitational acceleration.

From new experiments in a $0.4$m diameter column with ${\cal{O}}(10^3)$ particle Reynolds number bubbles and from a detailed analysis of published data, we confirm the self-organization prevailing in the heterogeneous regime, and that the liquid flow rate is only set by the column diameter $D$. Besides, direct liquid and gas velocity measurements demonstrate that the relative velocity increases above the terminal velocity $U_T$ in the heterogeneous regime, and that it tends to $\sim 2.4 U_T$ at very large gas superficial velocities $V_{sg}$. The proposed velocity scaling is shown to hold for liquid and gas mean velocities and for their standard deviations. Furthermore, it is found to be valid over a wide range of conditions, corresponding to Froude numbers $Fr=V_{sg}/(gD)^{1/2}$ from 0.02 to 0.5. Then, the relevance of this scaling for coalescing media is discussed. Moreover, following the successful prediction of the void fraction with a Zuber \& Findlay approach at the beginning of the heterogeneous regime, we show how the void fraction is correlated with $Fr$. Further investigations are finally suggested to connect the increase in relative velocity with meso-scale structures known to exist in the heterogeneous regime.

\end{abstract}

\begin{keywords}

\end{keywords}

\section{Introduction} \label{intro}

\NOTE{In bubble columns gas is injected at the bottom of an initially stagnant liquid contained in a vertical cylinder. Such systems are often used in industry as reactors (chemical and biochemical transformations), in separation techniques (flotation), to promote agitation and mixing (metallurgy)... At low inlet gas flow rate, bubbles homogeneously rise over the column cross-section. In consequence, mixing and turbulence are mainly generated by interactions of bubble wakes, and the void fraction is linearly increasing with the gas superficial velocity $V_{sg}$ ($V_{sg}$ is defined as the ratio of the injected gas flow rate to the entire column cross-section). When the inlet gas flow rate is increased above some threshold, the flow loses its spatial uniformity and its unsteadiness growths. In this so-called heterogeneous regime, complex flow structures appear as secondary motions are superimposed on the mean recirculation arising at the reactor scale. These motions are said to be `chaotic' by No\"el De Nevers who noticed in 1968: ``In unbaffled systems these circulations are unstable and chaotically change in size, shape, and orientation. These chaotic circulations provide the principal mode of vertical bubble transport in bubble columns over a wide range of operating conditions''. In addition, the increase of the void fraction with $V_{sg}$ turns to become non-linear (\cite{joshi1998gas,ruzicka2001effect,sharaf2016global}).

In industrial applications, superficial velocities are typically in the range 5 to 30 cm/s and the gas volume fraction evolves between 10 and 40\%: most of these systems are thus operated in the heterogeneous regime that corresponds to values of gas hold up above 20\%. A key issue in this regime is the extrapolation of results acquired in small scale experiments to full-size reactors those diameter $D$ can be as large as many meters. Indeed, in spite of a sustained scientific production over more than 70 years, the hydrodynamics of bubble columns operated in the heterogeneous regime is not yet fully understood. In particular, there is still no consensus on the scaling of key variables such as void fraction, mean liquid velocity, velocity fluctuations... As an illustration of that situation, more than twenty different correlations are currently proposed to evaluate the void fraction (see the reviews by \cite{deckwer1992bubble,joshi1998gas,kantarci2005bubble,rollbusch_bubble_2015,kikukawa_physical_2017,besagni_two-phase_2018}).} 

\NOTE{Similarly, and despite progresses in two-fluid modelling, simulations of bubble columns based on two-fluid approaches have not yet reached a fully predictive status (\cite{shu2019multiscale}). Indeed, ad-hoc adjustments are still required to reach some agreement with experiments. Concerning the momentum transferred from bubbles to the liquid, in earlier attempts it was the bubble size that was adjusted to modulate the relative velocity (\cite{ekambara2005cfd}). However, the current approach consists in multiplying the drag force by an ad-hoc coefficient function of the local void fraction in order to represent an effective momentum exchange between phases. That correction combines an increase of the drag at low void fractions - the well-known hindering effect (e.g. \cite{ishii1979drag} that was originally identified on solid suspensions by \cite{richardson1954sedimentation}), with a neat decrease at larger void fractions - with the so-called swarm effect (\cite{ishii1979drag,simonnet2007experimental}). This swarm effect represents the impact of neighbour bubbles on the motion of a test inclusion: it notably arises from the entrainment of bubbles in the wake of larger inclusions. In air/water systems, drag reduction has been observed to start at void fractions $\sim15\%$, and to reach a factor 5 at a void fraction of about 30\%. Various empirical expressions have been proposed for the swarm coefficient: let us quote \cite{roghair2011drag,mcclure2017experimental,gemello2018cfd,yang2018drag}... Despite slight variations (on the value of the critical void fraction, on some extra dependency on the mean bubble sizes), all these proposals are quite similar and they all mainly depend on the local void fraction. When used in bubble column simulations, such swarm correction leads to a strong increase of the apparent relative velocity in the heterogeneous regime, a trend that is in qualitative agreement with experiments (\cite{krishna1991model,raimundo2019hydrodynamics}). Provided that the bubble size is known beforehand (implying that coalescence/breakup mechanisms - whose modelling are also important issues - are not active in the flow conditions considered), such corrections ensure a reasonable agreement with air/water experiments with deviations up to 15\% on void fractions (global and local) and up to 40\% on the liquid velocity on the axis (\cite{ertekin2021validation}). Interesting, these figures hold over a significant range of conditions, namely for gas superficial velocities up to 25cm/s and for column diameters from 0.2 to 3m, indicating that the swarm effect is a key feature of the heterogeneous regime.} 

\NOTE{Others uncertainties in the modelling of bubble column hydrodynamics arise on the description of the turbulence in the liquid phase for which different approaches have been attempted (\cite{khan2017comparison}) and a new production model has been proposed (\cite{panicker2020computational}). This issue also concerns the description of the agitation induced by bubbles as a variety of mechanisms are able to generate velocity fluctuations in the continuous phase (\cite{risso2018agitation}) including cluster induced turbulence (\cite{capecelatro2015fluid,shu2019multiscale,panicker2020computational,baker2020reynolds}) due to meso-scale structures. Furthermore, the  presence of such structures in bubble columns operated in the heterogeneous regime has been demonstrated in recent experiments (\cite{raimundo2019hydrodynamics}).}

\NOTE{Hence, despite the limitations of current turbulence and agitation models, the introduction of an ad-hoc swarm coefficient in simulations happens to provide a somewhat reasonable agreement with experiments in terms of void fraction. Its seems thus that some robust physics takes place in the complex flows prevailing in the heterogeneous regime. To identify that physics, we revisit in this paper the hydrodynamics of bubble columns. Our starting point is} that recent experimental results support the idea that a strong analogy exists between a bubble column in the heterogeneous regime and turbulent buoyancy-driven flows in confined channels with zero mean flow. This prompted us to hypothesize that a dynamical equilibrium between inertia and buoyancy holds in the heterogeneous regime: such an equilibrium leads to a liquid velocity that scales as $V_{liquid} \sim \left( gD\varepsilon \right)^2$, where $g$ is the gravitational acceleration, $D$ the column diameter and $\varepsilon$ stands for the void fraction (\cite{cartellier2019bubble}). In this paper, that scaling proposal is analysed and tested against previous experimental data reported in the literature.
 \NOTE{ It is also tested against a new experimental dataset collected in a $D=0.4$m column that includes gas phase velocity statistics measured with a newly developed Doppler optical probe (\cite{lefebvre2019new,lefebvre2022new}).} We will notably show that, for bubble columns operated in the heterogeneous regime, both mean and fluctuating axial velocities of the liquid phase and of the gas phase closely follow the proposed scaling. Moreover, and to complement the velocities scalings, an empirical expression will be proposed for the void fraction that is backed up by a dimensional analysis, and by a Zuber \& Findlay (\cite{zuber1965average}) one-dimensional model relating void fraction and gas flow rate fraction. The comparison of that model with experiments shows that the axial evolution of the flow is significant, and that it deserves to be investigated to capture the flow structuration in the heterogeneous regime. 

\section{New velocity scaling proposal based on the equilibrium between inertia and buoyancy \label{sec2}}

Prior to the discussion on velocity scaling, \NOTE{let us first underline a few characteristics of hydrodynamics of bubble columns in the heterogeneous regime}. In a previous campaign (\cite{raimundo2019hydrodynamics}), controlled air-water experiments were performed over a wide range of column sizes (diameter $D$ from 0.15 to 3m) and superficial velocities ($V_{sg}$ from 3 to 35cm/s) while keeping a fixed bubble size. More precisely, the Sauter mean horizontal diameter was the same within the range $\pm1mm$ in all columns at a given value of $V_{sg}$: it increased from 6 to 8mm over the range of $V_{sg}$ studied with a mean eccentricity always close to 0.7. Accordingly, the equivalent diameter evolved from 4.7 to 7mm, and thus, the terminal velocity was nearly constant, equal to $0.21\pm0.01$cm/s whatever the flow conditions. Coalescence being avoided (or at least, it was too weak to influence the flow behavior), we have been able to clarify some key features of bubble columns operated in the heterogeneous regime. 

\NOTE{Notably}:

\begin{itemize}

\item The homogeneous-heterogeneous transition was observed with a fixed bubble size, meaning that, \NOTE{contrary to the current belief}, coalescence is not necessary for such a transition to occur.
 
\item \NOTE{Local void fraction fluctuations happen to be very significant in the heterogeneous regime: they evolve between one tenth to ten times the average gas hold-up when quantified by a one-dimensional Vorono\"i analysis (\cite{raimundo2015analyse,mezui2018characterization,raimundo2019hydrodynamics}).} 

\end{itemize}

The co-existence of `dense' regions corresponding to clusters of bubbles with regions almost `free of bubbles' called `voids' (\cite{raimundo2019hydrodynamics}) induces strong differences in local velocities: \NOTE{the bubble} transport is controlled by these clusters/voids meso-scale structures. \NOTE{This could be the origin of} the observed increase in the apparent relative velocity of the gas in the heterogeneous regime, an effect usually \NOTE{accounted by ways of} a swarm coefficient. \NOTE{Moreover,} the presence of clusters and voids in the mixture induces strong local shear rates as well as intense 3D vortical structures that are expected to significantly contribute to turbulence production. Clearly, the fluctuations in the mixture density induce strong spatial and temporal fluctuations in buoyancy (see figure \ref{fig:fig1} and videos in the supplementary material): they are thus reminiscent of convective instabilities arising in turbulent buoyancy driven flows \NOTE{with zero mean flow}.

\begin{figure}
\centering
\includegraphics[width=\textwidth]{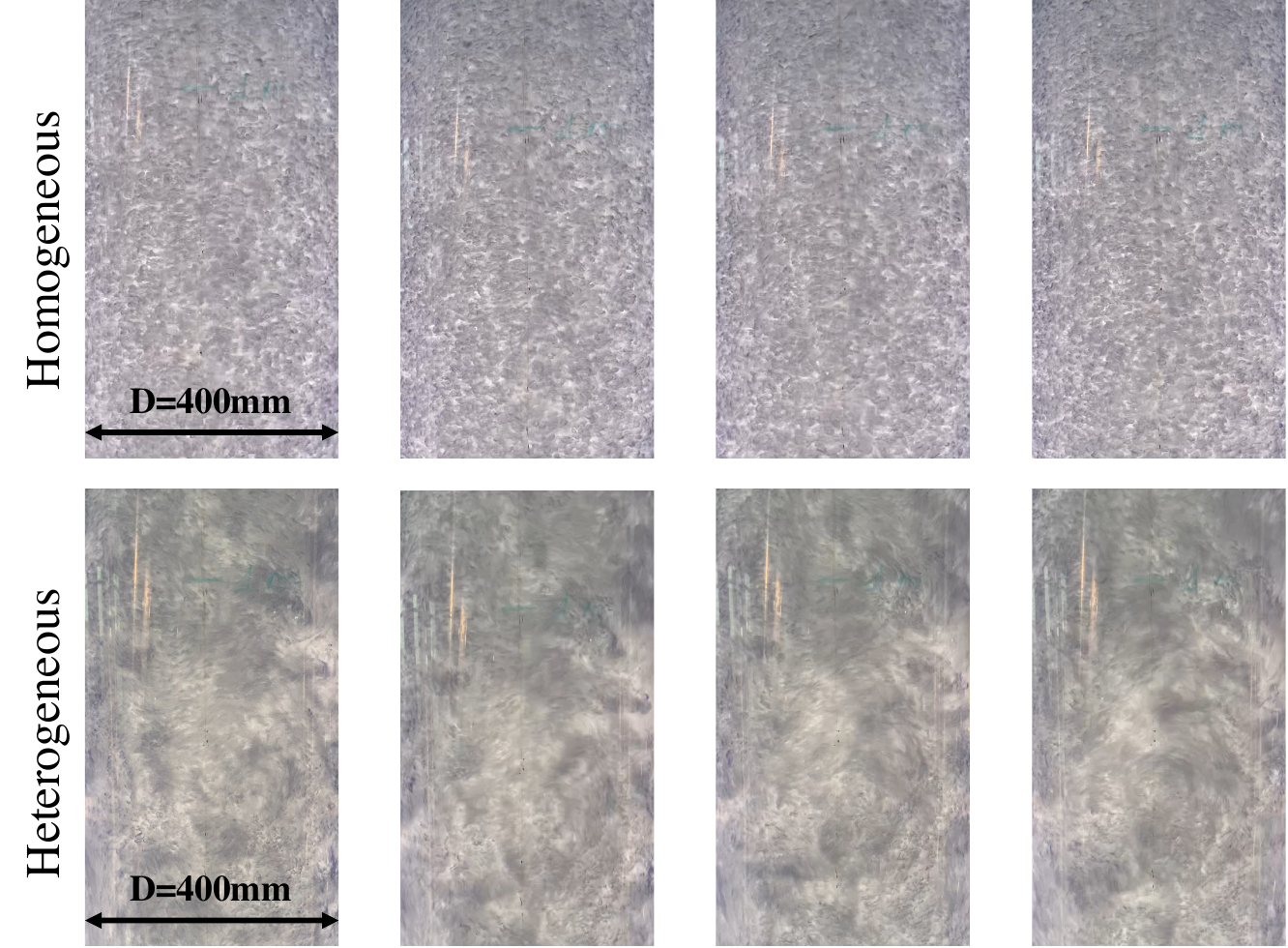}
\caption{\NOTE{Images of the flow in the vicinity of column walls between about 0.8 and 2 meters above gas injection, in the homogeneous regime (top row) and in the heterogeneous regime (bottom row). As side lightning was used, the grey level is an indication of the presence of bubbles: liquid structures comprising few bubbles appear as dark zones while clusters of bubbles correspond to bright zones. Results correspond to an air-water bubble column with $D=0.4m$ and static liquid height $H_0=2m$. The unsteadiness of these structures can be appreciated from the movies included in the supplementary material. The time increment between images is $1/30$s.}} \label{fig:fig1}
\end{figure}

Some analogies also hold on the global flow structure. Let us first discuss the existence of a quasi-fully developed region in bubble columns operated in the heterogeneous regime. Such a region appears in systems that are not too strongly confined: according to \cite{wilkinson1992design}, a minimum internal diameter of 0.15m is a necessary condition (possibly, that condition is helpful to avoid the development of a slug flow regime). In addition, the aspect ratio should be large enough so that end effects do not affect the flow organization in the central portion of the column: \cite{wilkinson1992design} argue that the dynamic height $H_D$ of the mixture should exceed 5 diameters while \cite{forret2003} established that $H_D/D=3$ is a sufficient condition for large (namely $D=0.4$m and 1m) columns. When introducing the static liquid height $H_0$, these conditions transform into $H_0/D \geq 3.8$ or 2.3, respectively, indicating that the bubble column should not be operated in the shallow water limit for a quasi-fully developed region to exist. Moreover, when the above conditions are fulfilled, the way gas injection is performed has no impact on the flow organization outside the entrance region. That conclusion has been ascertained in air-water systems with a gas injection evenly distributed over the column cross-section, and for large enough injection orifices (orifice diameter above 1mm according to \cite{wilkinson1992design}, above 0.5 mm according to \cite{sharaf2016global}). 

When one considers the so called `pure' heterogeneous regime, i.e. flow conditions such that the void fraction versus the superficial velocity is concave and that are far enough from the homogeneous/heterogeneous transition (\cite{ruzicka2013stability,sharaf2016global}), that conclusion seems to hold also when changing the coalescence efficiency by way of surfactants or of water purity (\cite{sharaf2016global}). The precise extend of the quasi-fully developed region is not exactly known: it is said to range from the end of the entrance region those extend is about one (\cite{forret2006scale}) or two (see \cite{guan2016experimental} and references therein) column diameter, up to typically one column diameter below the free-surface (\cite{forret2006scale}). Within that quasi-fully developed region, the self-similarity of the flow structure in the heterogeneous regime - in terms of transverse profiles of void fraction and liquid mean as well as fluctuating velocities - was shown to hold for diameters ranging from 0.15m up to 3m, and for superficial velocities spanning almost a decade, that is from the transition that arises for $V_{sg}$ about a few cm/s up to around $35-40$cm/s (\cite{forret2006scale,raimundo2019hydrodynamics}). 

Let us now consider the structure of the flow organisation. Globally, there is no imposed external pressure gradient. Instead, the buoyancy due to bubbles injected at the bottom forces the liquid upward. As there is no liquid flux entering or leaving the column, the liquid must flow downward in some regions of the system. In the homogeneous regime, that occurs in between bubbles everywhere in a given cross-section. In the heterogeneous regime, a stable global recirculation takes place with an upflow region in the centre and a downflow region near the walls. Indeed, the mean liquid velocity profiles consistently exhibit an inversion of the velocity direction at a distance from the column axis equal to $0.7R$ where $R$ is the column radius. 

Such large-scale organisation is reminiscent of turbulent buoyancy-driven flows in confined channels with zero mean flow where density gradients are due to solute concentration (\cite{cholemari2009axially}) or to temperature (\cite{castaing2017turbulent}). As for long channels or pipes, the translational invariance observed along the bubble column axis implies that the only characteristic length is the column diameter, provided that the column is high enough. Besides, and owing to that translational invariance, a uniform density gradient develops along the channel (\cite{cholemari2009axially,castaing2017turbulent}). That expectation is corroborated in bubble columns. Indeed, in the quasi-fully developed region, the local void fraction measured on the column axis $\varepsilon_{axis}$, exhibits a linear growth with the vertical distance above the gas injector, and the slope $d\varepsilon_{axis}/d(H/R)$ is proportional to the mean void fraction in the column (see figure 26 in \cite{lefebvre2022new}). Neglecting the density of the gas compared to that of the liquid (all the experiments discussed hereafter were performed at ambient pressure conditions), the local density $\rho$ of the mixture becomes $\rho \sim (1-\varepsilon) \rho_L$ (with $ \rho_L$ the density of the liquid phase). As the void fraction increases with the height above the injector, the density decreases with the height. Hence, the vertical stratification observed on the bubble column axis is stable. This contrasts with most investigations made on zero-mean flow buoyancy-driven turbulence in pipes.  For the later, the boundary conditions imposed at the top and bottom ends of the vertical pipe or channel are fixed temperature or solute concentration corresponding to unstable situations, that may lead to intermittent up and down flows (\cite{gibert2009heat,rusaouen2014laminar}. 

While in such studies the output is to evaluate the vertical flux of heat or of solute concentration, in bubble columns the boundary conditions are different since it is the gas volumetric flux through the system that is imposed, and the void fraction is the unknown. Owing to the local flow structure evoked above, injected bubbles are entrained in the central portion of the column, and most of them disappear at the free surface. A small fraction of these bubbles (about a few percent, see \cite{lefebvre2022new}) recirculate along the walls. These transport mechanisms lead thus to a strong transverse gradient in void fraction that induces radially distributed axial buoyancy forcing. Therefore, in bubble columns, the flow destabilization arises from the radial distribution of the two phases (instead of an unstable vertical stratification). 

In the liquid momentum balance, fluid inertia terms (namely $\rho \partial_t v_{Li}$, $\rho v_{Lj}\partial_jv_{Li}$ and $\partial_ip$) equilibrate buoyancy. Within a Boussinesq approximation, the later equals $g_i\Delta\rho$ where $g$ is the gravitation acceleration and $\Delta\rho$ the difference in density at the origin of buoyancy forces. Hence, the velocity scale for the liquid obeys:

\begin{equation}\label{c7eq1}
V_L \sim \sqrt{(gD\Delta\rho/\rho)},
\end{equation}

\noindent where $\Delta\rho/\rho$ is evaluated at a large length scale. In turbulent buoyancy-driven flows in confined channels, the constant axial gradient of the mean density is used to evaluate $\Delta\rho$ (\cite{cholemari2009axially,castaing2017turbulent}). In bubble columns, as the flow destabilization arises from lateral differences in density and thus in buoyancy, we sought a relevant scale from the radial void fraction profile. The void fraction typically evolves between $\varepsilon_{axis}$ on the column axis and nearly 0 in the wall zone. Thus, the magnitude of the radial difference in density $\Delta\rho$ over a length scale of the order of the column diameter $D$ is $\Delta \rho \sim D \partial \rho / \partial r \sim \rho_L D \partial \varepsilon / \partial r \sim \rho_L \varepsilon_{axis}$. Therefore, the void fraction on the axis $\varepsilon_{axis}$ is a measure of the radial density gradient $\Delta\rho/\rho_L$. \NOTE{The self-similarity of the radial void fraction profiles $\varepsilon/\varepsilon_{axis}=f(r/R)$ also supports that result.} \NOTEN{Such self-similarity is found empirically and therefore the functional form of $f(r/R)$ has not been deduced from first principles yet. Nevertheless, different fits of $f(r/R)$ have been proposed in the literature (see \cite{forret2006scale} and discussions therein).} 

Hence, as $\varepsilon_{axis}$ is proportional to the global void fraction (\cite{raimundo2016new}), one can use any characteristic gas fraction $\varepsilon$ in the system to estimate the magnitude of the radial density gradient  $\Delta\rho/\rho_L$. Consequently, the scaling from equation \ref{c7eq1} becomes:

\begin{equation}\label{eqscal}
V_L \sim \sqrt{(gD\varepsilon)}.
\end{equation}

The Boussinesq approximation is not mandatory for the derivation of Eq. \ref{eqscal}. Indeed, for a bubbly flow, the dynamical equilibrium for the liquid phase balances at first order inertia terms (i.e.  $\rho_L DV_L/Dt$) with the momentum transfer between phases. The later, homogeneous to a force per cubic meter, can be estimated as the void fraction  times the force $F$ exerted by a single bubble on the fluid divided by the bubble volume $\mathcal{V}$. Hence, the momentum equation for the liquid writes at first order:

\begin{equation}\label{eqscal1}
\rho_L DV_L/Dt=-\nabla{P}+\mu_L \Delta V_L+\varepsilon F / \mathcal{V},
\end{equation}

\noindent where the pressure gradient term $p$ includes the hydrostatic contribution. \NOTE{The last term in eq. \ref{eqscal1} represents the momentum source for the liquid phase due to the presence of bubbles: it is written here with the force $-F$ on a bubble because of the neat scale separation between the disturbance fields (evolving over at the scale close to the bubble size), and the undisturbed fields $V_L$ and $P$ (that change over a scale of order $D$). Given the dynamic equilibrium of the dispersed phase, and since there is no mean vertical acceleration of the continuous phase in the quasi-fully developed region,} the force $F$ along a vertical corresponds to the buoyancy on a bubble, i.e. to $\rho_L g_i \mathcal{V}$. The momentum transfer amounts then to $\varepsilon \rho_L g_i$, and the Eq. \ref{eqscal} scaling is recovered by balancing inertia and buoyancy without using the Boussinesq approximation,  that is without constraints on the relative velocity between phases.

Former experimental results support the velocity scaling proposed in Eq. \ref{eqscal}. Indeed, in \cite{raimundo2019hydrodynamics}, we evaluated the liquid flow rate $Q_{Lup}$ in the core region of the flow that is in the zone where the mean liquid velocity is upward directed. Owing to the self-organization of the flow occurring in the heterogeneous regime, that region extends from the column axis up to a radial distance of $0.7R$ (that $0.7R$ limit was also identified by \cite{kawase1986liquid}). We have shown that, in the heterogeneous regime, $Q_{Lup}$ is independent of the gas superficial velocity. Instead, $Q_{Lup}$ only depends on the column diameter $D$ and it scales as $D^{5/2}$: consequently, $(gD)^{1/2}$ was identified as the proper velocity scale for the mean flow circulation. Equation \ref{eqscal} is consistent with that result since the void fraction is known to be weakly sensitive (if any) to the column diameter. In the following, we test the relevance of Eq. \ref{eqscal} for bubble columns in the heterogeneous regime by examining a number of experimental results relative to the liquid and to the gas velocities, including mean and fluctuating components. In section \ref{sec3}, we consider new experiments in which we succeeded to gather reliable statistics on bubble velocity. In section \ref{sec4}, we examine data sets extracted from the literature.

\section{Test of the scaling on new gas and liquid velocity data collected in a $D=0.4m$ bubble column \label{sec3}}


A new optical probe that combines accurate phase detection (its sensing length is very small, equal to $6\mu$m) with gas velocity measurements based on Doppler signals collected from approaching interfaces has been recently developed based on a technology patented by A2 Photonic Sensors. The probe design, the signal processing and the sensor qualification are detailed in \cite{lefebvre2022new}, where mean bubble velocity profiles in a $D=0.4$m bubble column are also presented. In the following, we exploit further that probe to examine how bubble velocity statistics evolve with the gas superficial velocity. In parallel, classical Pavlov tubes are also used to access the liquid velocity. Let us first summarise the experimental conditions.

\subsection{Experimental conditions}

The experiment consisted in a $3$m high and $D=0.4$m internal diameter bubble column functioning with air and water. The gas injector was a 10mm thick plexiglass plate perforated by 352 orifices of 1mm internal diameter. These orifices were uniformly distributed over the column cross-section. The column was filled with tap water at an initial height $H_0=2.02$m. The surface tension of the tap water used was $67$mN/m at $25^\circ$C, its pH evolved in the interval $[7.7, 7.9]$ and its conductivity varied within the range $330-450 \mu$S/cm, indicating the presence of a significant solid content. All the data presented here were gathered at $H=1.45$m above injection, that is at $H/D=3.625$, a position well within the quasi fully developed region. Besides, and owing to the large ratio $H_0/D=5.05$, the information collected in that zone is not sensitive to the static liquid height $H_0$. Experiments were performed for superficial gas velocities $V_{sg}$ ranging from 0.6cm/s to 26cm/s; \NOTE{the maximum global volume fraction was about 35\%}. 

Information relative to bubbles was acquired with the Doppler probe. For each bubble detected, the probe gives access to the gas residence time $t_G$, that is the time spent by the probe tip inside the bubble. \NOTE{The void fraction is given by the sum of residence times divided by the measuring duration. For the present flow conditions in terms of fluids, bubble size, absolute velocity of bubbles and probe dimension, the uncertainty on void fraction measurements is less than 1\% according to the study of \cite{vejravzka2010measurement} in air-water systems.} Besides, a Doppler signal is recorded from the rear interface (that is at the gas to liquid transition) for some bubble signatures, and its analysis provides the bubble velocity $V_b$ projected along the fiber axis. When both the gas residence time $t_{Gi}$ and the bubble velocity $V_{bi}$ are available for the ith bubble, one can infer the gas chord $C_i = V_{bi} t_{Gi}$ cut by the probe through that bubble. \NOTE{The typical reproducibility on chord length measurements is of order 3\% on the mean value, and less than 20\% on the standard deviation (\cite{lefebvre2022new}).}

For the liquid phase, velocity statistics were measured with a Pavlov device made of two parallel tubes (external diameter 6mm, internal diameter 5mm), each drilled with a 0.5mm in diameter hole. These two orifices faced opposite directions: they were aligned along a vertical, and the vertical distance between them was 12mm. The pressure transducer was a Rosemount 2051 CD2 with a dynamics of $\pm15000$ Pa, a resolution of $\pm9.75$ Pa and a response time of $130$ms. The differential pressure was transformed into the local liquid velocity using $v_L^2(t) = \pm 2 \vert p(t) \vert / \rho_L$ (no correction dependent on void fraction was considered) with the appropriate sign. Hence, the dynamics in velocity was $\pm5.48$m/s and the resolution $\pm0.14$m/s.

\begin{figure}
     \centering
     \begin{subfigure}[b]{0.3\textwidth}
         \centering
         \includegraphics[width=\textwidth]{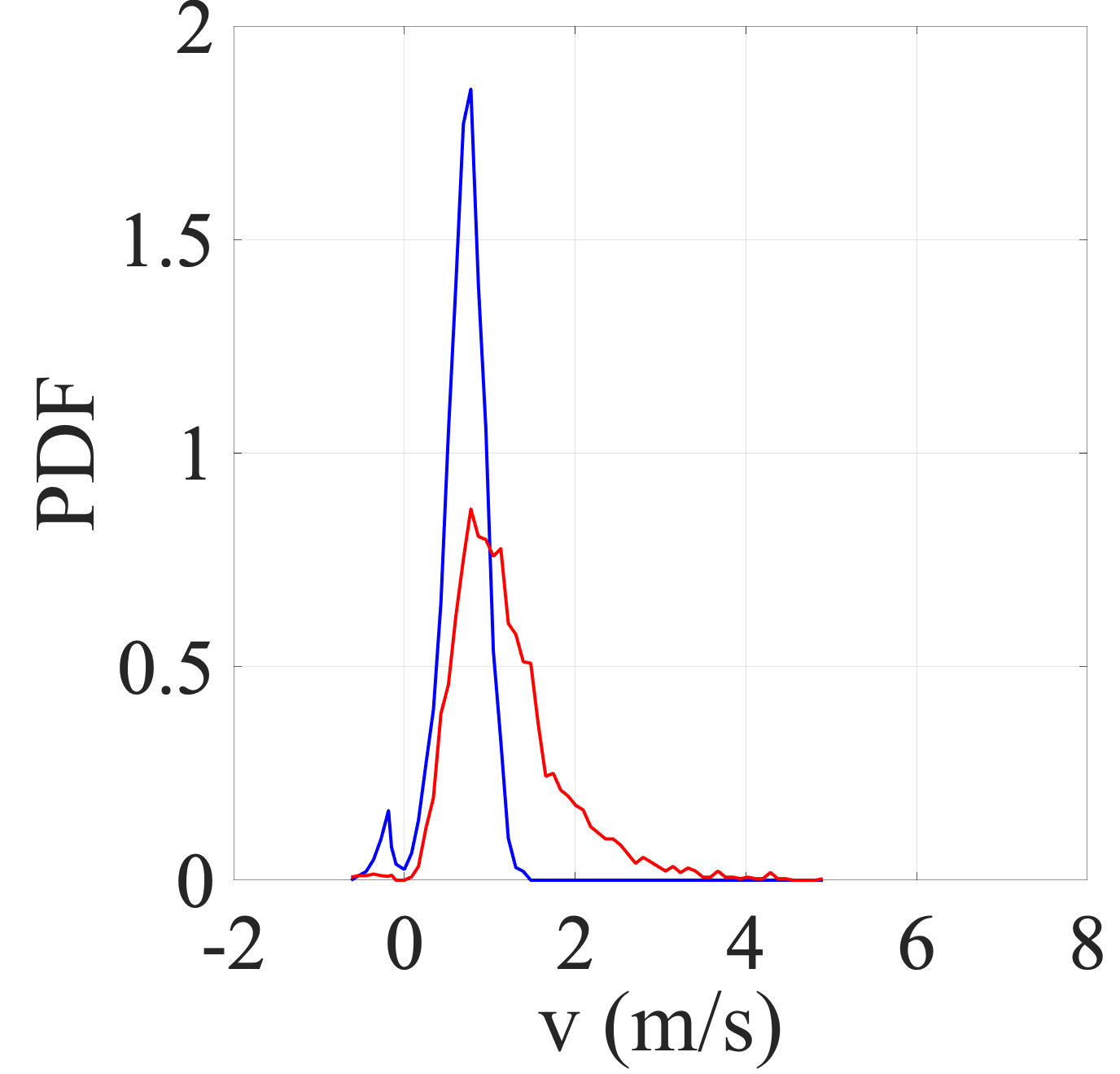}
         \caption{}
         \label{fig2a}
     \end{subfigure}
     \hfill
     \begin{subfigure}[b]{0.3\textwidth}
         \centering
         \includegraphics[width=\textwidth]{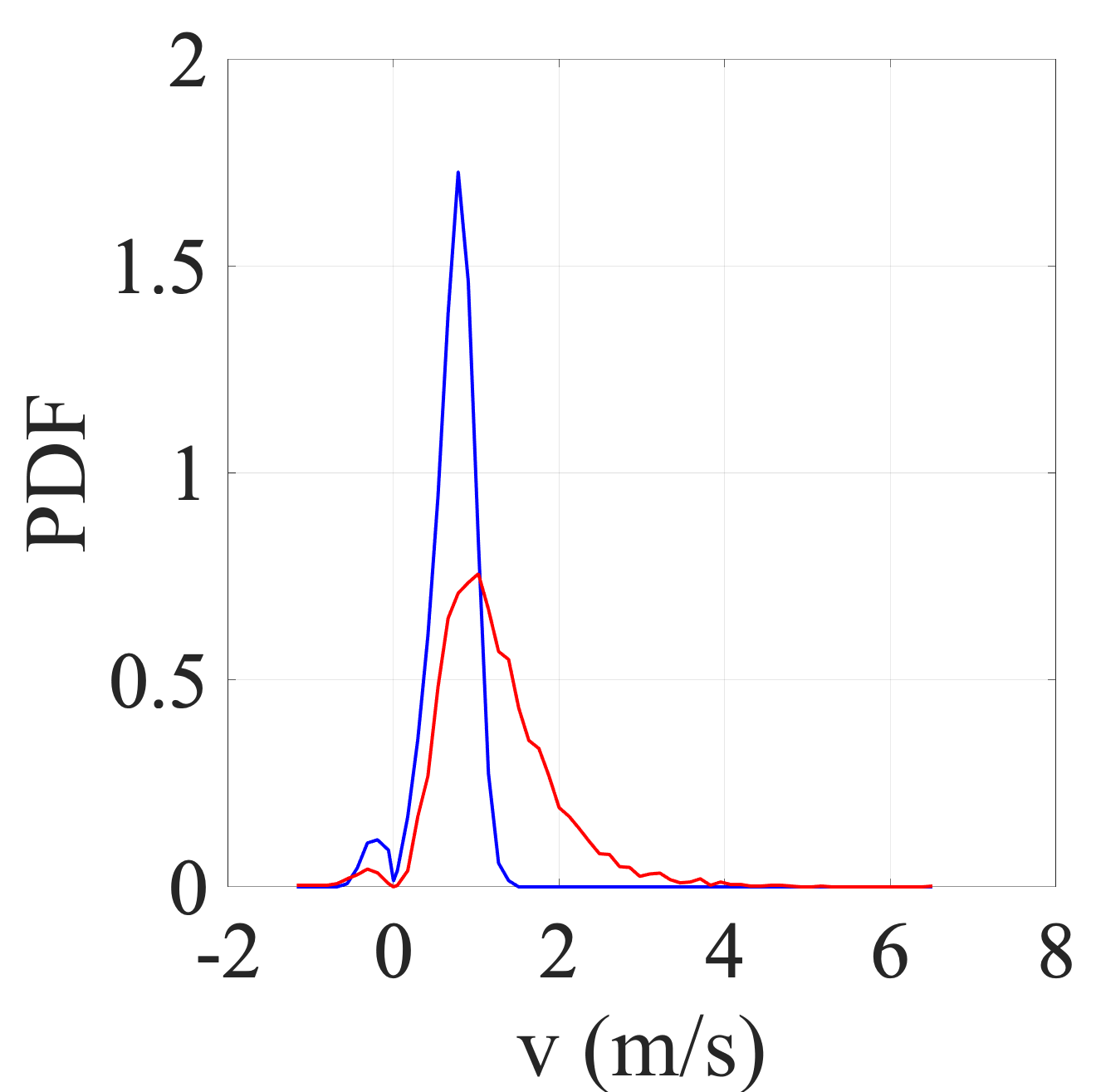}
         \caption{}
         \label{fig2b}
     \end{subfigure}
     \hfill
     \begin{subfigure}[b]{0.3\textwidth}
         \centering
         \includegraphics[width=\textwidth]{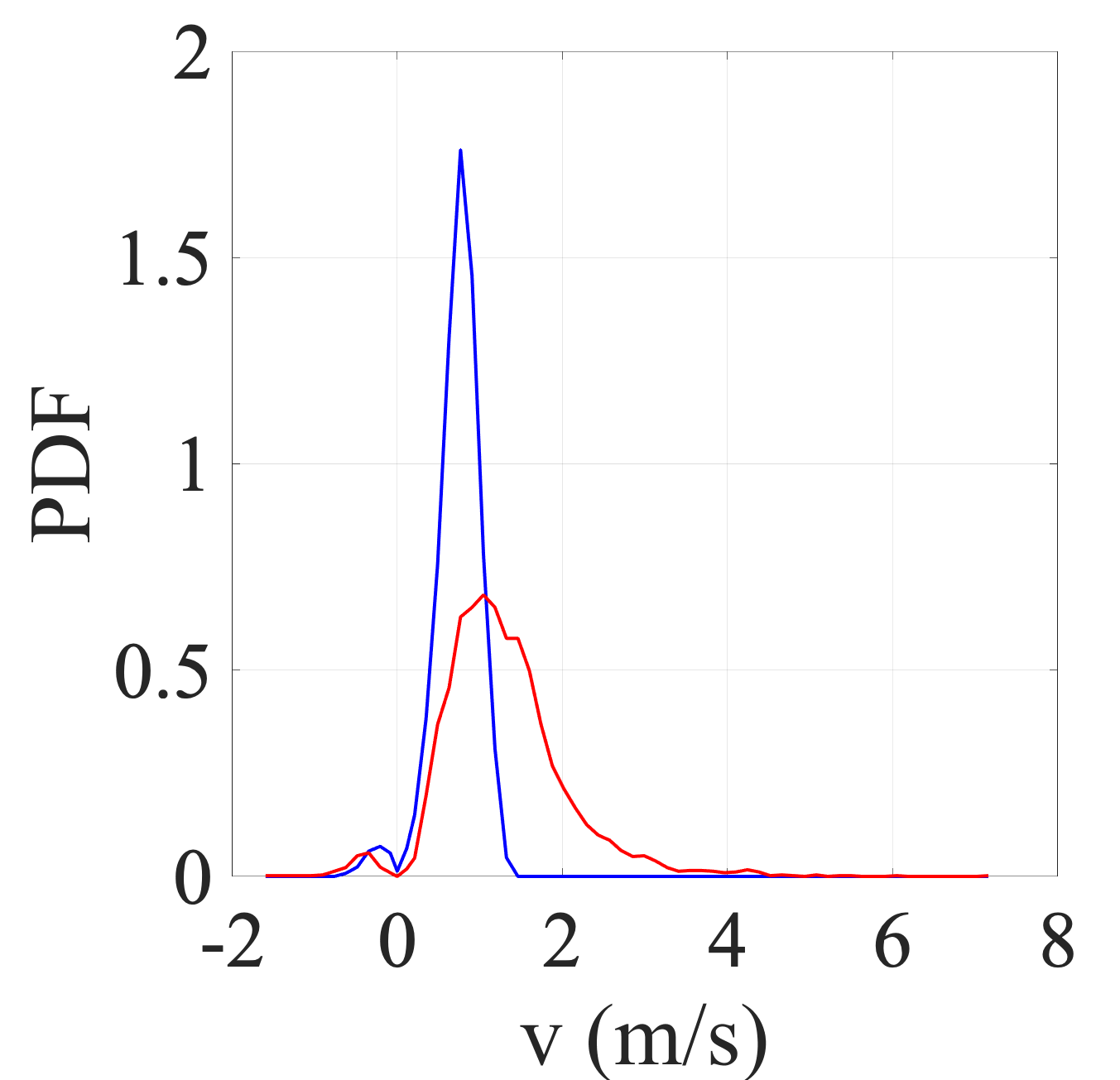}
         \caption{}
         \label{fig2c}
     \end{subfigure}
        \caption{Velocity pdfs for the liquid (blue lines) and for the bubbles (red lines) measured on the column axis at $H/D=3.625$ for $v_{sg}=13$cm/s (a), $v_{sg}=16.25$cm/s (b) and $v_{sg}=22.75$cm/s (c).}
        \label{fig2}
\end{figure}

Liquid and gas velocity pdfs measured with these sensors in the center of the column are illustrated Figure \ref{fig2}. By construction, the Pavlov tube detects both positive (upward directed, i.e. against gravity) and negative (downward directed, i.e. along gravity) velocities. For bubbles, as the Doppler probe detects only inclusions approaching it head on, the pdfs were built by cumulating the information gathered over the same measuring duration and at the same position with an upward directed probe and with a downward directed probe. More details and discussion concerning these measurements are presented in \cite{lefebvre2022new}. \NOTE{The reproducibility on mean bubble velocity and on its standard deviation is better than $\pm 5\%$.}

Concerning the bubble size, the analysis of the axial evolution of chords distributions along the column indicates that coalescence was absent (or at least extremely weak) in our experimental conditions (\cite{lefebvre2022new}), most probably because of the partial contamination of the tap water used. Over the investigated range of superficial velocities, the Sauter mean vertical diameter of bubbles remained in the interval $[6.2mm; 6.7mm]$ while their Sauter mean horizontal diameter measured with the correlation technique (\cite{raimundo2016new}) increased with $V_{sg}$ from 6.6 to 7.8mm. Overall, the mean equivalent bubble diameter remained in the interval $[6.62mm; 7.35mm]$: that corresponds to a terminal velocity from 21cm/s to 23cm/s (\cite{maxworthy1996experiments}) and to particle Reynolds numbers in the range $1450-1550$. 

\begin{figure}
     \centering
     \begin{subfigure}[b]{0.45\textwidth}
         \centering
         \includegraphics[width=\textwidth]{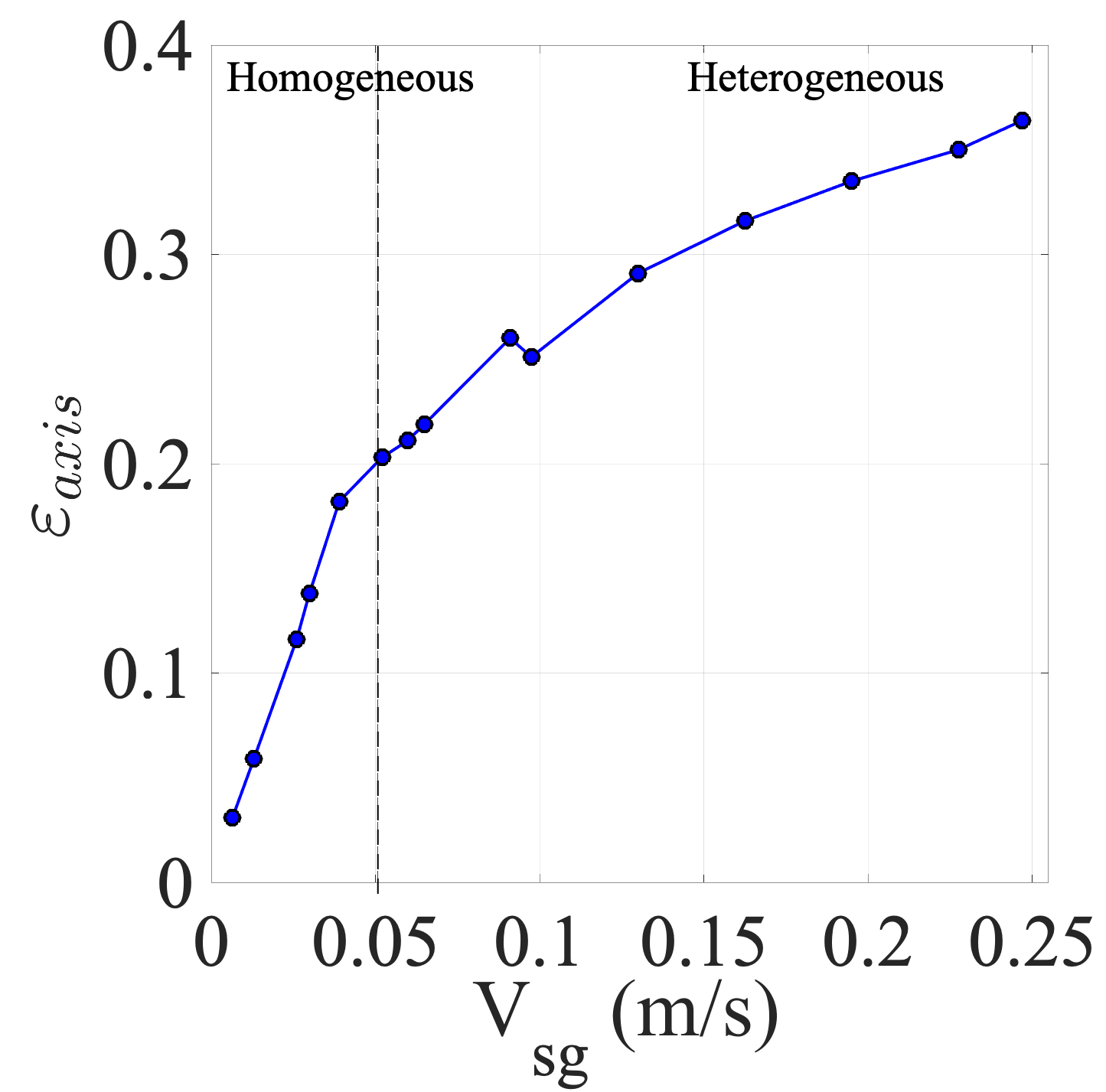}
         \caption{}
         \label{fig3a}
     \end{subfigure}
     \hfill
     \begin{subfigure}[b]{0.47\textwidth}
         \centering
         \includegraphics[width=\textwidth]{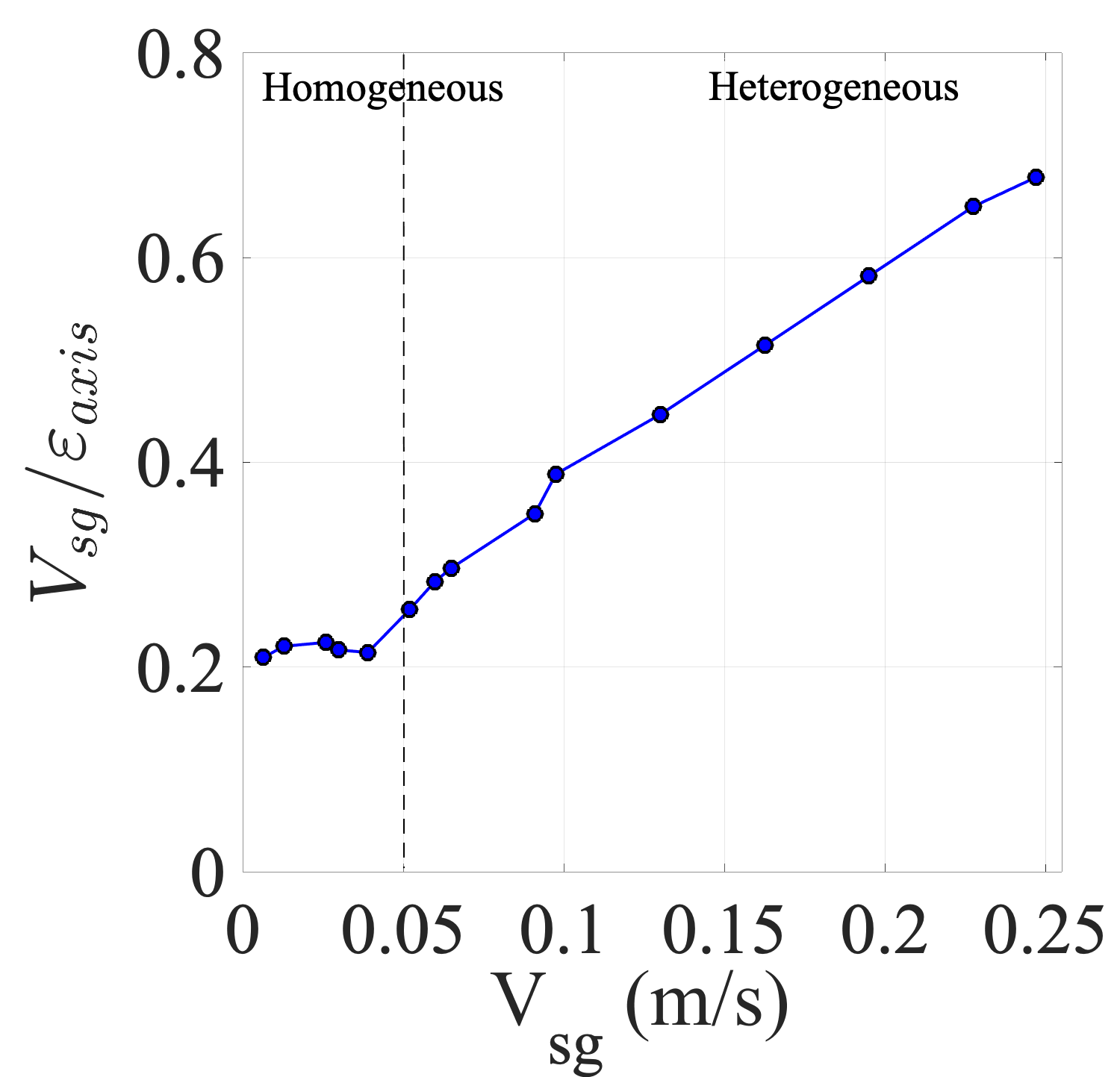}
         \caption{}
         \label{fig3c}
     \end{subfigure}
        \caption{(a) Evolution of the local void fraction $\varepsilon_{axis}$ on the column axis with the gas superficial velocity $V_{sg}$. (b) Plot of the apparent rise velocity estimated as $V_{sg}/\varepsilon_{axis}$ versus $V_{sg}$. Measurements with a downward directed Doppler probe at an height $H/D=3.625$ above injection.}
        \label{fig3}
\end{figure}

The local void fraction measured on the column axis $\varepsilon_{axis}$ is plotted versus the gas superficial velocity in Figure \ref{fig3}a. The homogeneous regime ends for $V_{sg}$ between $\sim4$cm/s and $\sim5$cm/s, while the `pure' heterogeneous regime starts at $V_{sg} \sim 6.5cm/s$. Following \cite{krishna1991model}, these data are plotted as $V_{sg}/\varepsilon_{axis}$ versus $V_{sg}$ in Figure \ref{fig3}b: they exhibit a constant rise velocity, close to the bubble terminal velocity $U_T$, up to $V_{sg} \sim 4$cm/s, that is within the homogeneous regime. Beyond that, the apparent rise velocity (called `rise velocity of swarm' by \cite{krishna1991model}), monotonously increases with the gas superficial velocity. It reaches a magnitude of about 3 times $U_T$ at the largest $V_{sg}$ investigated here (namely 24.7cm/s). That increase is the signature of the heterogeneous regime. Note also that the latter correspond to a void fraction on the axis that exceeds 20\%. In the following, the transition will be represented by a vertical dash line at $V_{sg}=5$cm/s in figures as a guide to the eye. 

\subsection{Local gas and liquid velocity on the column axis versus $V_{sg}$ \label{sec32}}

\begin{figure}
\centering
\includegraphics[width=\textwidth]{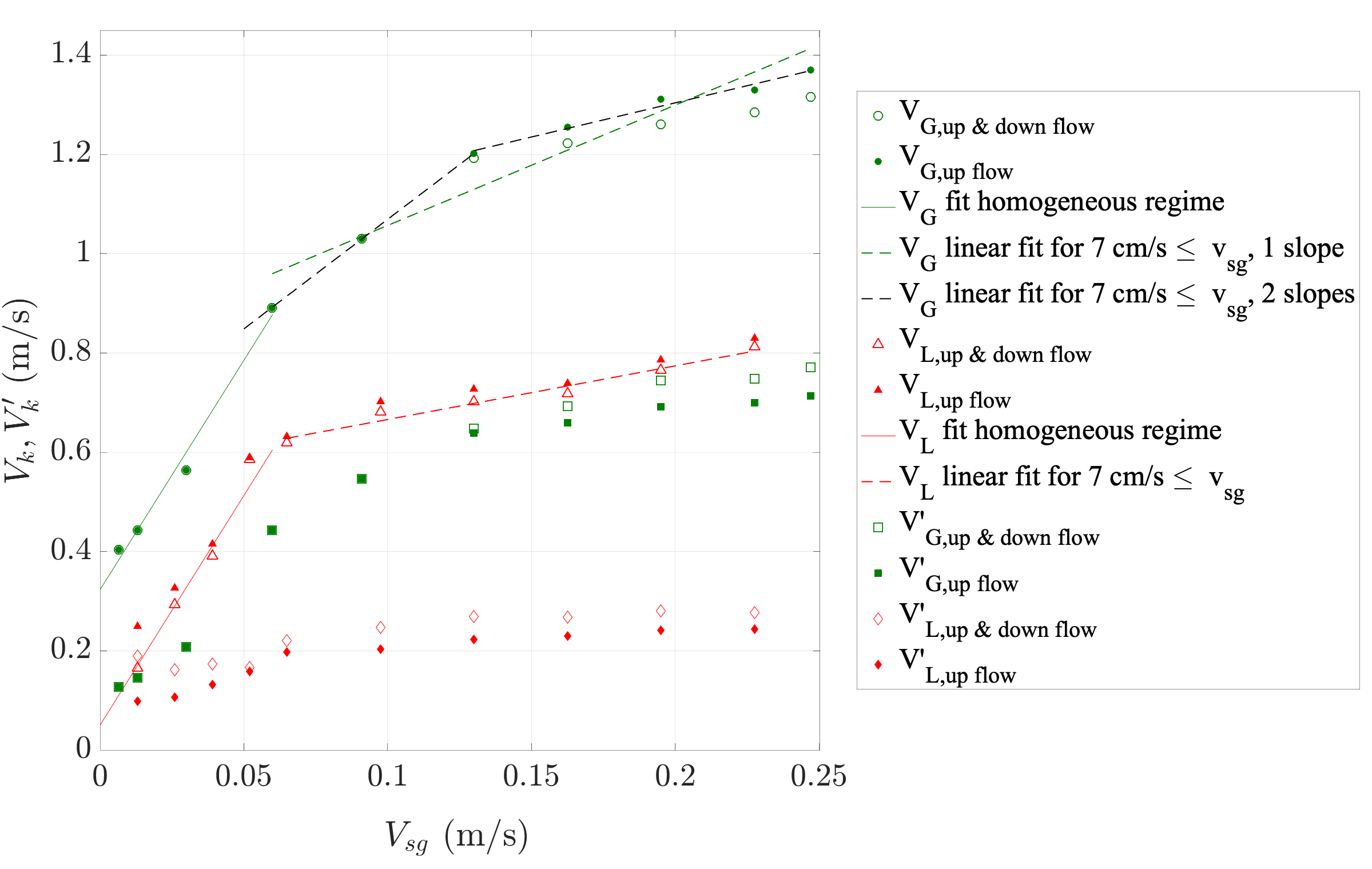}
\caption{Evolution of the mean vertical velocities of the bubbles $V_G$ and of the liquid $V_L$, and of their standard deviation ($V_G'$ for the gas, $V_L'$ for the liquid), with the gas superficial velocity $V_{sg}$. Measurements performed in a $D=0.4$m column, at $H/D=3.625$ and on the column axis. The bubble velocities were measured with a Doppler probe and the liquid velocities with a Pavlov tube. The straight lines in the homogeneous (continuous lines) and in the heterogeneous (dashed lines) regimes are linear fits of the data. Note that in the heterogeneous regime, two plausible trends (green and black dashed lines) are proposed for the mean bubble velocity. The difference between `up flow' and `up and down flow' sets is explained in the text. } \label{fig4}
\end{figure}

Figure \ref{fig4} provides the mean vertical velocities of bubbles $V_G$ and of the liquid $V_L$ on the column axis as well as the standard deviations $V_G'$ for the gas phase and $V_L'$ for the liquid phase. Two datasets are presented for the mean velocities. 

For bubbles, a set named `up flow' corresponds to measurements achieved with a Doppler probe pointing downwards that collects only upward directed (i.e. positive) vertical velocities. A second dataset named `up and down flow' was obtained by gathering direct (i.e. without interpolation) velocity measurements from a probe pointing downward, with direct (i.e. without interpolation) velocity measurements from the same probe pointing upward. In this process, the measuring duration was the same for the two probe orientations. In the flow conditions considered here, the mean velocities from these two sets are close, with a difference of at most 4\% (\cite{lefebvre2022new}). Similarly, the difference on bubble velocity standard deviations from these two sets is at most 8\%. 

For the liquid velocities, two datasets are also presented: one corresponds to moments evaluated over the entire distribution (named `up and down flow') while the other concerns positive velocities only (named `up flow'). In the heterogeneous regime, the difference between the two sets is at most 3.6\% for the mean value and 18\% for the standard deviation. Oddly, larger liquid velocity deviations between `up flow' and `up and down flow' statistics appear in the homogeneous regime. The difference is especially pronounced for $V_{sg}$ below 3cm/s. These deviations are related with the unexpected apparition of a significant negative tail in the liquid velocity pdfs when $V_{sg}$ becomes small, a defect that may possibly be due to the flow perturbation induced by the rather large probe holder used in our experimental setup. 

Except for these low $V_{sg}$ cases, the differences between the `up flow' and the `up and down flow' datasets remain weak. These small differences partly come from the fact that these data are all collected on the column axis, where the probability of occurrence of absolute negative velocities in the laboratory frame remains small. Indeed, in our experiments, the probability to observe a downward directed liquid velocity on the axis was less than 3\% for any $V_{sg}$ in the heterogeneous regime. Similarly, \cite{xue2008bubble} found a probability for observing negative bubble velocities on the column axis between 4\% and 4.5\% at $V_{sg}=14$cm/s and about 6\% at $V_{sg}=60$cm/s. For the gas phase, and because of the positive (upward directed) relative velocity, these probabilities should be lower than the above figures. Hence, considering either `up flow' or `up and down flow' data sets does not change the conclusions proposed hereafter. Yet, the distinction between the two series is worth to be kept in mind in the perspective of analyzing other radial positions where the probability of occurrence of downward flow increases. 

\begin{figure}
\centering
\includegraphics[width=\textwidth]{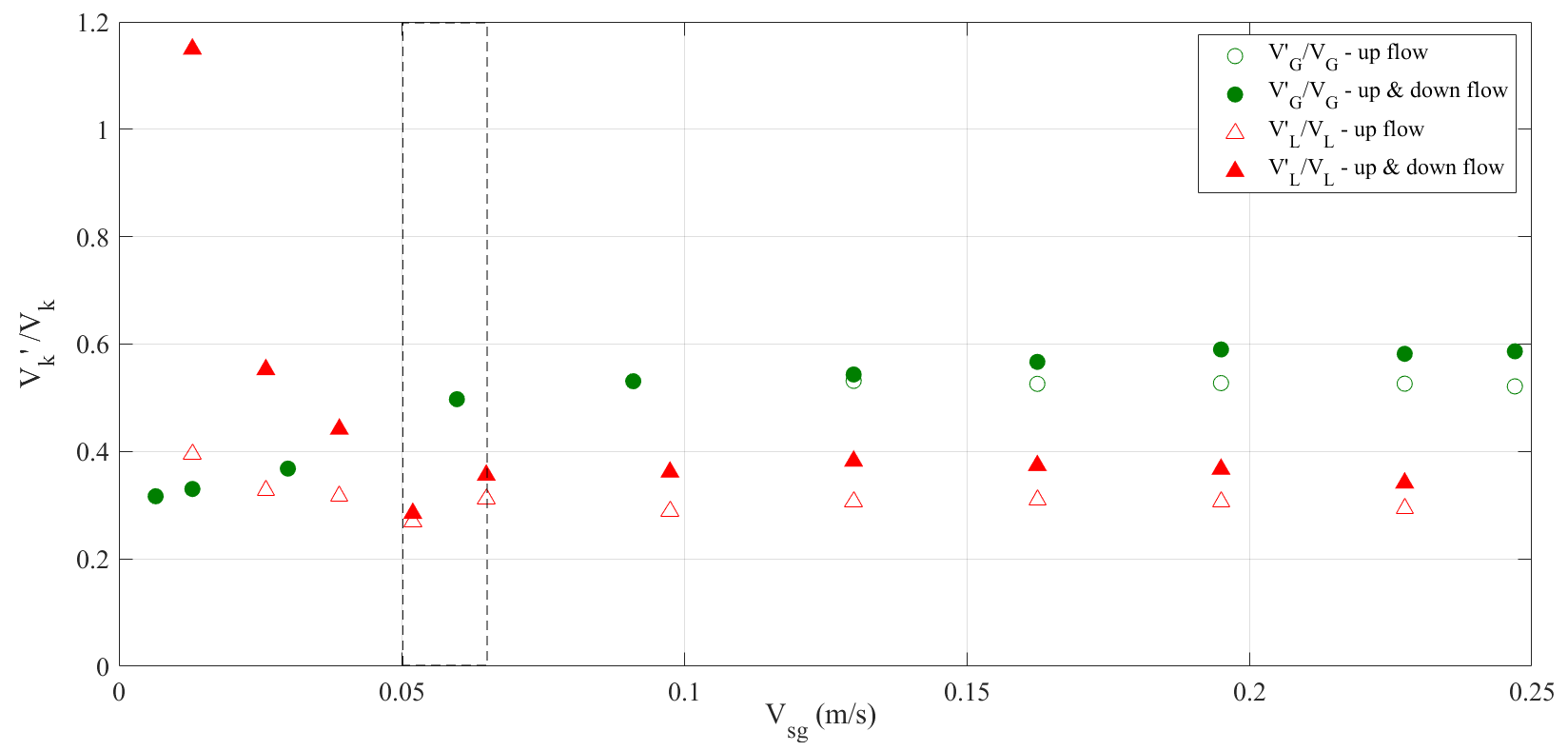}
\caption{Vertical velocity fluctuations $V'/V$ of liquid and gas phases versus the gas superficial velocity $V_{sg}$. Measurements performed in a $D=0.4$m column, on the column axis at $H/D=3.625$.} \label{fig5}
\end{figure}

From the data presented Figure \ref{fig4}, a series of comments and conclusions emerge:

\begin{enumerate}

\item the local relative velocity $V_G-V_L$ remains nearly constant in the homogeneous regime. It is about 27cm/s, a magnitude comparable to the `mean' terminal velocity $U_T$ of the bubbles generated in the column. 

\item in the heterogeneous regime, the relative velocity becomes larger than $U_T$. Furhermore, it seems to monotonously increase with $V_{sg}$ (see the trend indicated by the green and red dashed lines in Figure \ref{fig4}). At about $V_{sg}=19.5$m/s, the measured relative velocity amounts to 54.6 cm/s that is 2.5 times the terminal velocity $U_T$. Thus, these direct velocity measurements are consistent with the behavior of the `rise velocity of swarm' $V_{sg}/\varepsilon_{axis}$ shown in Figure \ref{fig3}b. They also confirm the conclusion we previously obtained (see \cite{raimundo2019hydrodynamics}) by analyzing the flow in the core region of the bubble column: the apparent relative velocity was indeed found to range between 2 and 8 times $U_T$ using a conservative evaluation of the gas liquid flow rate in the core region of the bubble column.

\item the relative fluctuations in velocity $V'/V$ are nearly constant in the heterogeneous regime (as shown Figure \ref{fig5}). For the liquid phase, the average of $V_L'/V_L$ over the data collected in the heterogeneous regime equals 36.5\%, in agreement with previous findings (see the discussion in \cite{raimundo2019hydrodynamics}). For the gas phase, the average of $V_G'/V_G$ is even higher; it equals 52-53\% when considering positive velocities only, and it rises up to 56-57\% when combining positive and negative velocities measurements (\cite{lefebvre2022new}). These strong figures confirm that intense turbulent motions take place in heterogeneous conditions. 

\item from a closer examination of Figures \ref{fig4} and \ref{fig5}, two different behaviors could possibly be distinguished in the heterogeneous regime. From the homogeneous/heterogeneous transition up to $V_{sg}$ about $13-15$cm/s, the relative velocity $V_G-V_L$ and also to some extend the fluctuation $V_G'/V_G$ exhibit a clear monotonous increase with the gas superficial velocity. Above $V_{sg} \sim 13-15$cm/s, these two quantities seem to become constant. In particular, the increase in the bubble velocity illustrated by the black dashed line in Figure \ref{fig4} becomes nearly parallel to that of the mean liquid velocity (dashed red line in Figure \ref{fig4}): accordingly, the relative velocity seems to stabilize at a value about $2.3-2.4 U_T$ at large $V_{sg}$. With regard to flow dynamics and scaling laws, it would be worthwhile to clarify whether the relative velocity reaches an asymptote, or if it continues to grow with $V_{sg}$: more experimental data covering an enlarged range of gas superficial velocities are required for that. 
\end{enumerate}

\begin{figure}
\centering
\includegraphics[width=\textwidth]{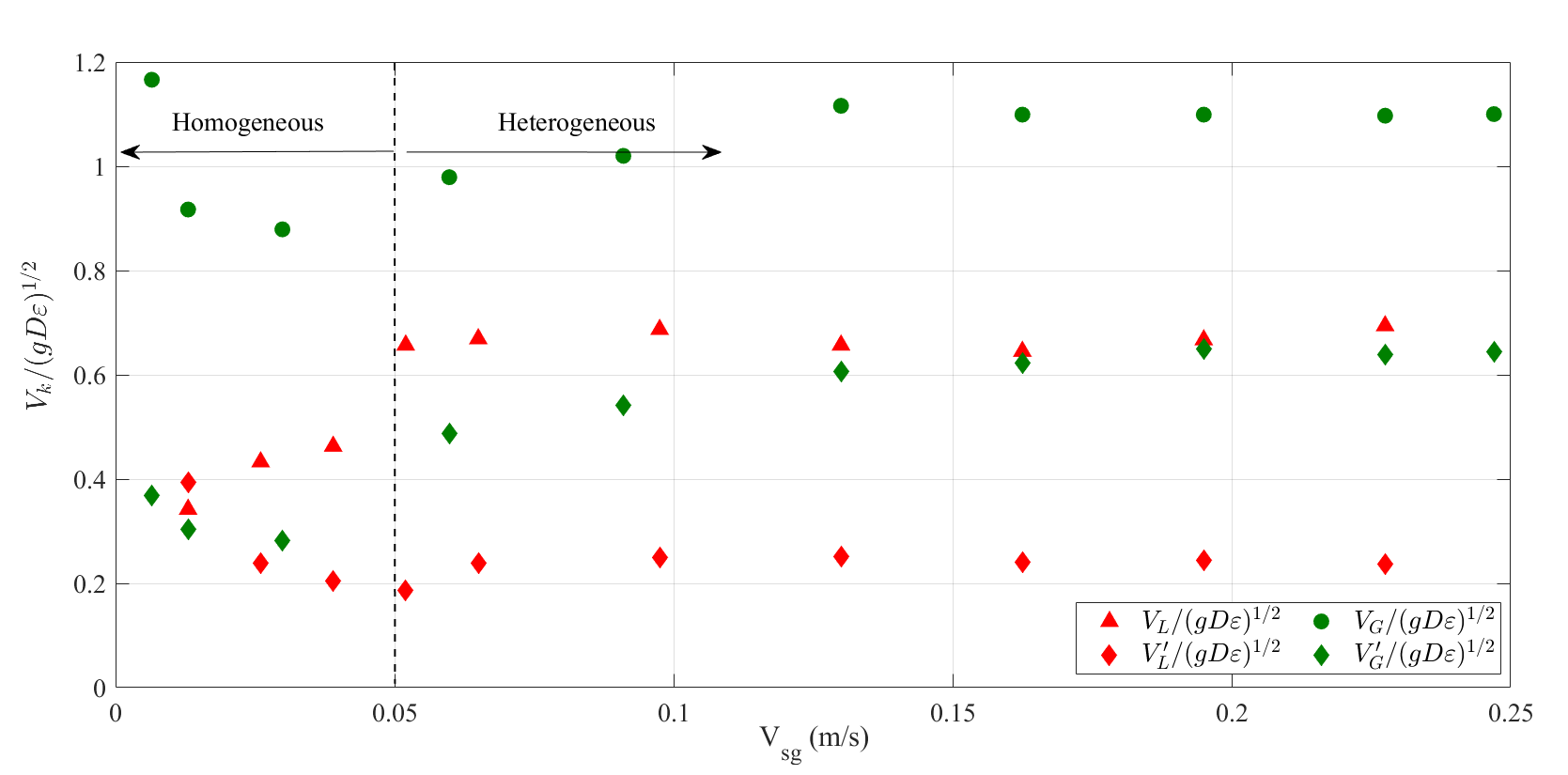}
\caption{Evolution of phasic velocities (`up and down flow' velocity statistics) scaled by $(gD\varepsilon)^{1/2}$ versus the superficial velocity $V_{sg}$. The mean ($V$) and fluctuating ($V'$) components of the bubble and the liquid vertical velocities as well as the void fraction $\varepsilon$ are local quantities measured in a $D=0.4$m column, on the column axis at $H/D=3.625$.} \label{fig6}
\end{figure}

To test the scaling proposed in eq. \ref{eqscal} on these experimental data, we used local void fraction and velocities measured on the axis and at the same height $H/D$ above the injector. Since the two sets `up flow' and `up and down flow' are close, let us consider only `up and down flow' velocity statistics for the analysis. The mean velocities as well as the standard deviations scaled by $(gD\varepsilon)^{1/2}$ are represented for both phases in Figure \ref{fig6}. Clearly, all these quantities remain nearly constant for all flow conditions pertaining to the heterogeneous regime. For the mean bubble velocity on the column axis, one gets

\begin{equation}\label{eq4}
V_G/(gD\varepsilon)^{1/2} \sim 1.09\pm0.02.,
\end{equation}

\noindent while for the mean liquid velocity on the column axis:

\begin{equation}\label{eq5}
V_L/(gD\varepsilon)^{1/2} \sim 0.68\pm0.01.
\end{equation}

According to these results, the relative velocity on the column axis scales as $U_R \sim 0.41 (gD\varepsilon)^{1/2}$, i.e. it monotonously increases with the void fraction $\varepsilon$. Such an increase of the relative velocity with the void fraction has been identified in bubble columns using indirect arguments (see for example \cite{raimundo2019hydrodynamics}). It has also been observed in others gas-liquid systems. Such a behavior is sometimes represented by a swarm coefficient that quantifies the decrease of the drag force acting on a bubble with the void fraction (\cite{ishii1979drag,simonnet2007experimental}). Nowadays, ad-hoc swarm coefficients are routinely introduced in numerical simulations based on two-fluid models (e.g. \cite{mcclure2017experimental,gemello2018cfd}). We bring here direct experimental evidence of the increase of the relative velocity with the void fraction in a bubble column operated in the heterogeneous regime. 

Concerning the standard deviation of velocities, the standard deviation being proportional to the mean (see Figure \ref{fig4}), they also follow the same scaling with $V_G'/(gD\varepsilon)^{1/2} \sim 0.6\pm0.02$ and $V_L'/(gD\varepsilon)^{1/2} \sim 0.22 \pm 0.02$. Hence, all the above results obtained on the mean and on the fluctuating components of bubbles and liquid velocities confirm the soundness of the scaling proposed in eq. \ref{eqscal} with respect to void fraction.  

The scaling resulting from an inertia-buoyancy equilibrium proposed above also includes an increase of velocities with the square root of the bubble column diameter. As the experiments presented in this section concern only one bubble column diameter, the dependency with the column diameter cannot be tested. New experimental data gathered in bubble columns of variable diameters are needed to test the validity of the proposed scale. In that perspective, available experiments from the literature and relative to bubble columns of variable diameter will be analyzed in the next section. 

Nevertheless, we already have accumulated strong experimental evidence of the relevance of the square root of the bubble column diameter as a scaling factor for the mean liquid velocity (\cite{raimundo2019hydrodynamics}). Indeed, the neat upward liquid flow rate $Q_{Lup}$ in the column, where $Q_{Lup}$ is obtained by integrating the liquid flux $(1-\varepsilon) V_L$ over the core region (i.e. from the axis up to $0.7-0.71 R$), happens not to depend on $V_{sg}$ in the heterogeneous regime. In the present experiments, we also found that $Q_{Lup}$ is independent on $V_{sg}$ in the heterogeneous regime. That conclusion was confirmed from data collected at three heights above injection, namely $H/D=2.625$, $H/D=3.625$ and $H/D=4.875$. The result is $Q_{Lup}=0.0122$ m$3$/s, with variations between $+0.001$ m$^3$/s and $0.0007$ m$^3$/s depending on the set of data considered to evaluate the mean. 

\begin{figure}
     \centering
     \begin{subfigure}[b]{0.49\textwidth}
         \centering
         \includegraphics[width=\textwidth]{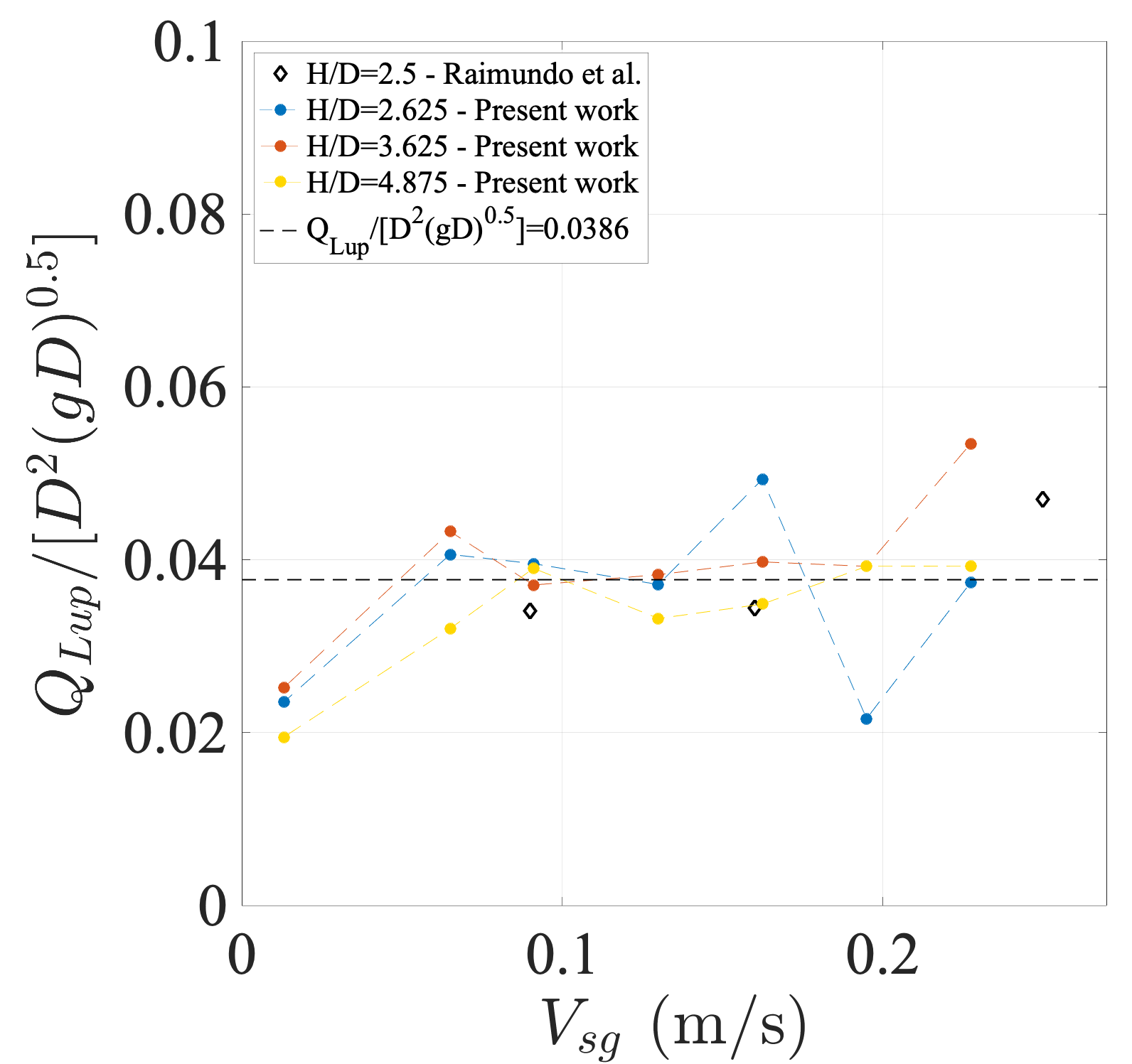}
         \caption{}
         \label{fig7a}
     \end{subfigure}
     \hfill
     \begin{subfigure}[b]{0.49\textwidth}
         \centering
         \includegraphics[width=\textwidth]{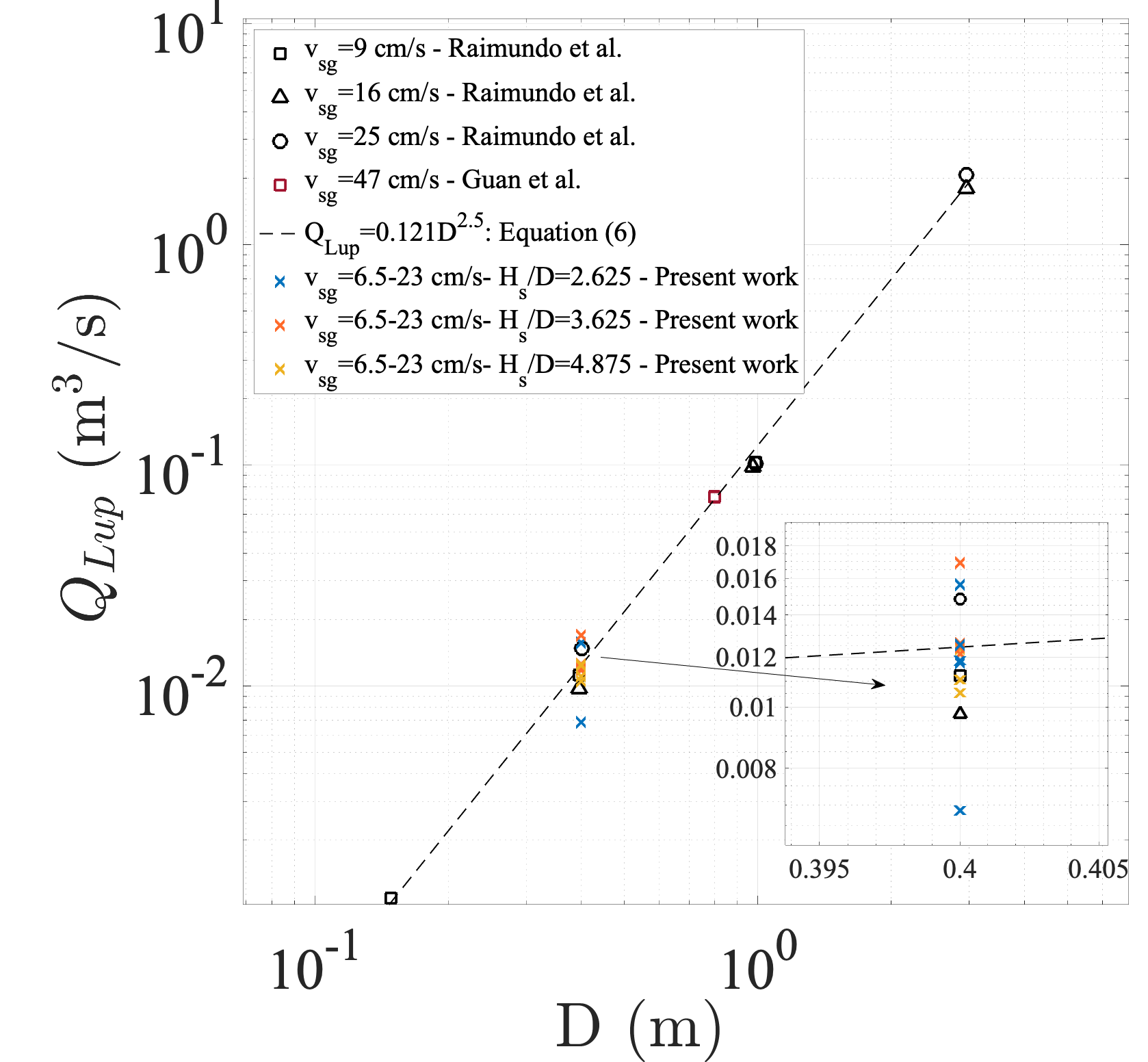}
         \caption{}
         \label{fig7b}
     \end{subfigure}
        \caption{(a) Upward directed liquid flux $Q_{Lup} / [D^2 (gD)^{1/2}]$ versus $Vsg$ measured at different heights above injection in $D=0.4$m columns. (b) Evolution of $Q_{Lup}$ with the bubble column diameter from \cite{raimundo2019hydrodynamics}, from \cite{guan2015hydrodynamics} and from present data. }
        \label{fig7}
\end{figure}

Moreover, we have previously shown (\cite{raimundo2019hydrodynamics}) that $Q_{Lup}$ is proportional to $D^2 (gD)^{1/2}$ over a significant range of column diameters (from $D=0.15$m to 3m) and of gas superficial velocities (from $V_{sg}=9$ to 25cm/s). As shown Fig. \ref{fig7}a, the present experiments confirm that finding in a $D=0.4$m column, for $V_{sg}$ between 6.5cm/s and 22.7cm/s and for $2.625 \leq H/D \leq 4.875$. The dependency of $Q_{Lup}$ with $D$ is further illustrated in Fig. \ref{fig7}b where we have reported the present data, the data collected by \cite{raimundo2019hydrodynamics}, as well as one data produced by \cite{guan2015hydrodynamics} in the following flow condition: $V_{sg}=47$cm/s in a $D=0.8$m column and for $2.75 \leq H/D \leq 4.625$. All these data fall onto the same curve. Overall, the observed liquid flow rate - column diameter relationship writes:

\begin{equation}\label{eq6}
Q_{Lup}=0.0386\pm0.002 D^2 (gD)^{1/2}.
\end{equation}

This result can also be expressed as a Froude number based on the average liquid velocity $Q_{Lup}/S_{core}$ in the core region. Here, the cross section $S_{core}$ of the ascending flow zone is evaluated as $\pi D^2/8$ (the mean liquid becomes zero at a radial position comprised between $0.7$ and $0.71R$: that limit is well approximated as $2^{1/2}/2 R= 0.707 R$). Hence, $Fr_L = (Q_{Lup}/S_{core}) / (gD)^{1/2} = 0.098$ \footnote{There is a typo error in eq.(13) of \cite{raimundo2019hydrodynamics}: the coefficient 0.024 should be replaced by 0.098.} . 
All the above mentioned experiments bring a clear evidence that the velocity of the mean liquid circulation in a bubble column operated in the heterogeneous regime scales as $(gD)^{1/2}$ with the column diameter.

\section{Confrontation of the proposed scaling with experimental data from literature \label{sec4}}

In this section, we examine whether available experimental data on local liquid or gas/bubble velocities obey the scaling proposed above when considering a larger range of flow conditions, and in particular variable column diameters. To this end, we target meaningful experiments in bubble columns in the sense that we seek conditions such that the flow dynamics is controlled by the same mechanisms as those discussed in section \ref{sec2}. More precisely, we select experiments pertaining to the `pure' heterogeneous regime, a regime that occurs well beyond the transition region, and for which the global void fraction is an increasing and concave function of the gas superficial velocity. In addition, we consider only data collected in the quasi-fully developed region of the bubble column: as discussed in section \ref{sec2}, that condition implies a minimum column diameter, a minimum static liquid height as well as a proper range of measuring heights. 

Besides, and although it is known that in the heterogeneous regime the flow organization is weakly sensitive to the injector design, we also select experiments such that the gas injection is pretty well uniformly distributed over the column cross-section to avoid forcing of large-scale instabilities by an uneven gas repartition at injection and/or to avoid strong asymmetry of the mean flow such as the one observed by \cite{chen2003local} in their largest column. 

Another constraint in the search of relevant data is that both local velocity and local void fraction should be simultaneously available. In the following, we focus on local data gathered on the column axis. The sets of data fulfilling the above-mentioned constraints are presented in tables \ref{tab1a}\&\ref{tab1b} for the liquid phase and in tables \ref{tab2a} \& \ref{tab2b} for the gas phase. Note that almost all experimental conditions in tables \ref{tab1a}, \ref{tab1b}, \ref{tab2a} and \ref{tab2b} correspond to air-water systems and to `large' bubbles, i.e. with an equivalent diameter between 3mm and 10mm. Their terminal velocity typically ranges between 21cm/s and 27cm/s (\cite{maxworthy1996experiments}), so that all these flow conditions involve bubble dynamics at high (from 800 to 2100) particle Reynolds number. 

Yet, these experiments remain difficult to quantify with respect to coalescence. We qualitatively estimated the coalescence efficiency based on measurements of the axial evolution of the bubble size when such data were available. When such information was absent, we considered how the bubble size changes with the superficial velocity: a strong increase of the latter from homogeneous to heterogeneous conditions could be (but this is not certain) the mark of a neat coalescence. In addition, let us underline that when a significant coalescence is present, the flow regime may continuously evolve with the height above injection so that a quasi-fully developed region may not exist or may require column heights much larger than the ones available in standard experiments. All the information collected is summarized in tables \ref{tab1a}, \ref{tab1b}, \ref{tab2a} and \ref{tab2b}, where the situation with respect to coalescence has been classified into three main categories: none or weak coalescence, medium coalescence, strong coalescence, and the situation is said unclear when information was insufficient to conclude.

\subsection{Mean liquid velocity on the column axis \label{sec41}}

\NOTEN{We start this section by briefly detailing the datasets from the literature that will be discussed.} Tables \ref{tab1a}\&\ref{tab1b} lists the experiments that obey the above constraints and that provide the liquid velocity and the local void fraction on the bubble column axis. Some choices were made to exploit the data. For \cite{Hills1974}, we considered only the data acquired with the `plate B' that corresponds to a uniform gas injection over the column cross section. In \cite{forret2006scale}, the only quantitative data on gas hold-up are global void fractions deduced from static and dynamic liquid heights. We transformed the global void fraction into a local void fraction on the axis by multiplying it by 1.5 as done by the authors (see their eq.(4)), but this factor could be inappropriate. The local void fractions for the data of \cite{vial2001influence} were collected in \cite{camarasa1999influence}. For \cite{yao1991bubble}, we present the data they collected at various heights with $H/D$ from 2.6 up to 12, and we use an extrapolation to evaluate the void fraction at $V_{sg}=10$cm/s. Note also that these authors imposed a forced liquid motion but the mean velocity of 1cm/s is negligible compared with the measured liquid and gas velocities on the column axis: these data are therefore believed to be representative of a bubble column operated in a batch mode and they have been kept in the analysis. Finally, all the measurements mentioned in tables \ref{tab1a}\&\ref{tab1b} considered positive and negative velocity realizations, although some (hard to evaluate) bias may be present notably with Pavlov tubes, as evoked by \cite{Hills1974}. To be consistent, we compared with our data series named `up and down flow' (see Section \ref{sec3}).

\begin{sidewaystable}%
  \vspace{16cm}
    \caption{\NOTEN{List of references and flow conditions exploited to extract liquid velocity and local void fraction on the column axis. Further information is provided in table \ref{tab1b}.}}
  \vspace{-1.5cm}

  \includegraphics[width=0.99\textwidth]{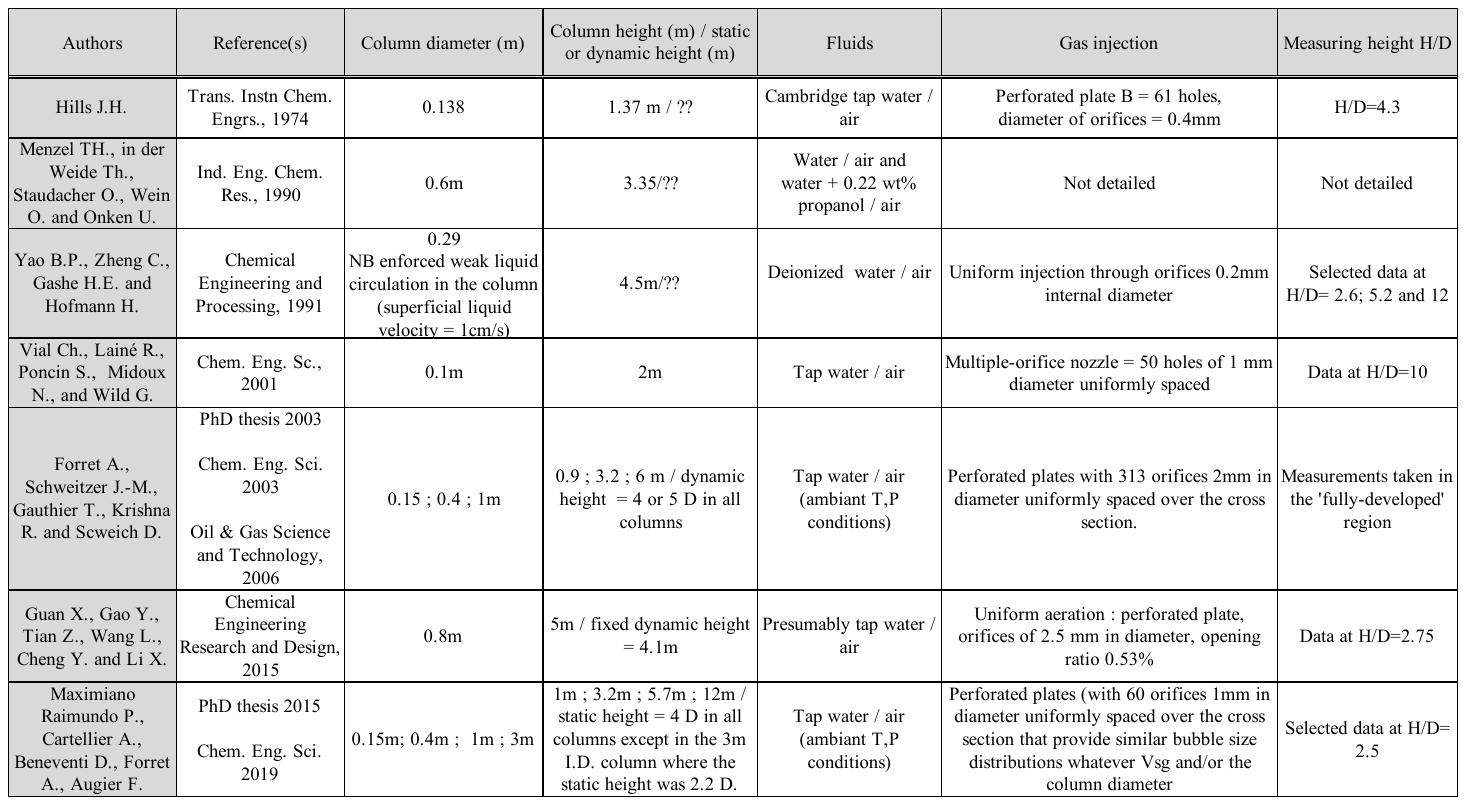}
\label{tab1a}
\end{sidewaystable}

\begin{sidewaystable}%
  \vspace{16cm}
  \caption{\NOTEN{List of references and flow conditions exploited to extract liquid velocity and local void fraction on the column axis (this table complements the information given in table \ref{tab2a}).}}
  \vspace{-1.5cm}  
  \includegraphics[width=0.99\textwidth]{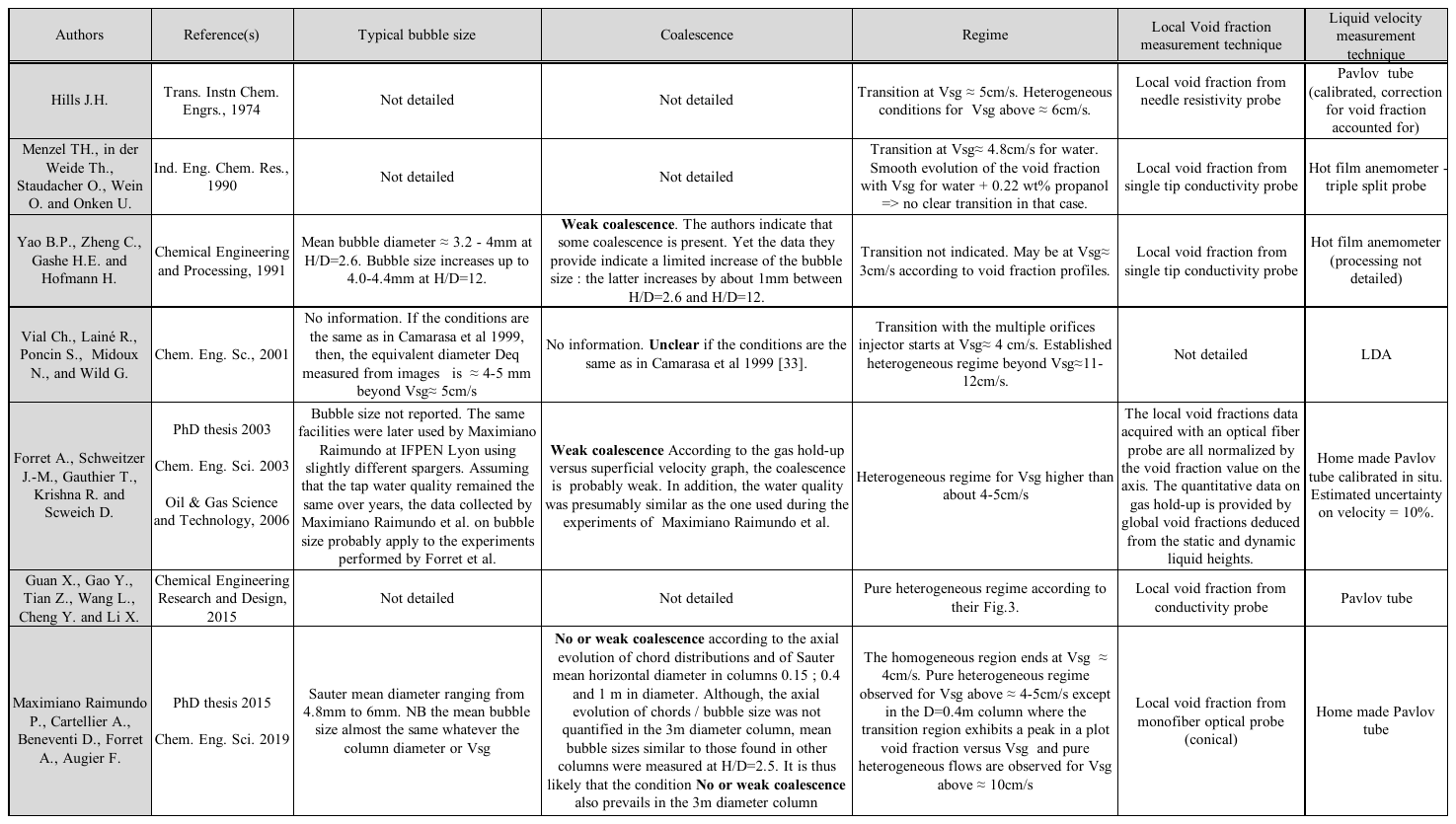}
\label{tab1b}
\end{sidewaystable}

\begin{figure}
\centering
\includegraphics[width=\textwidth]{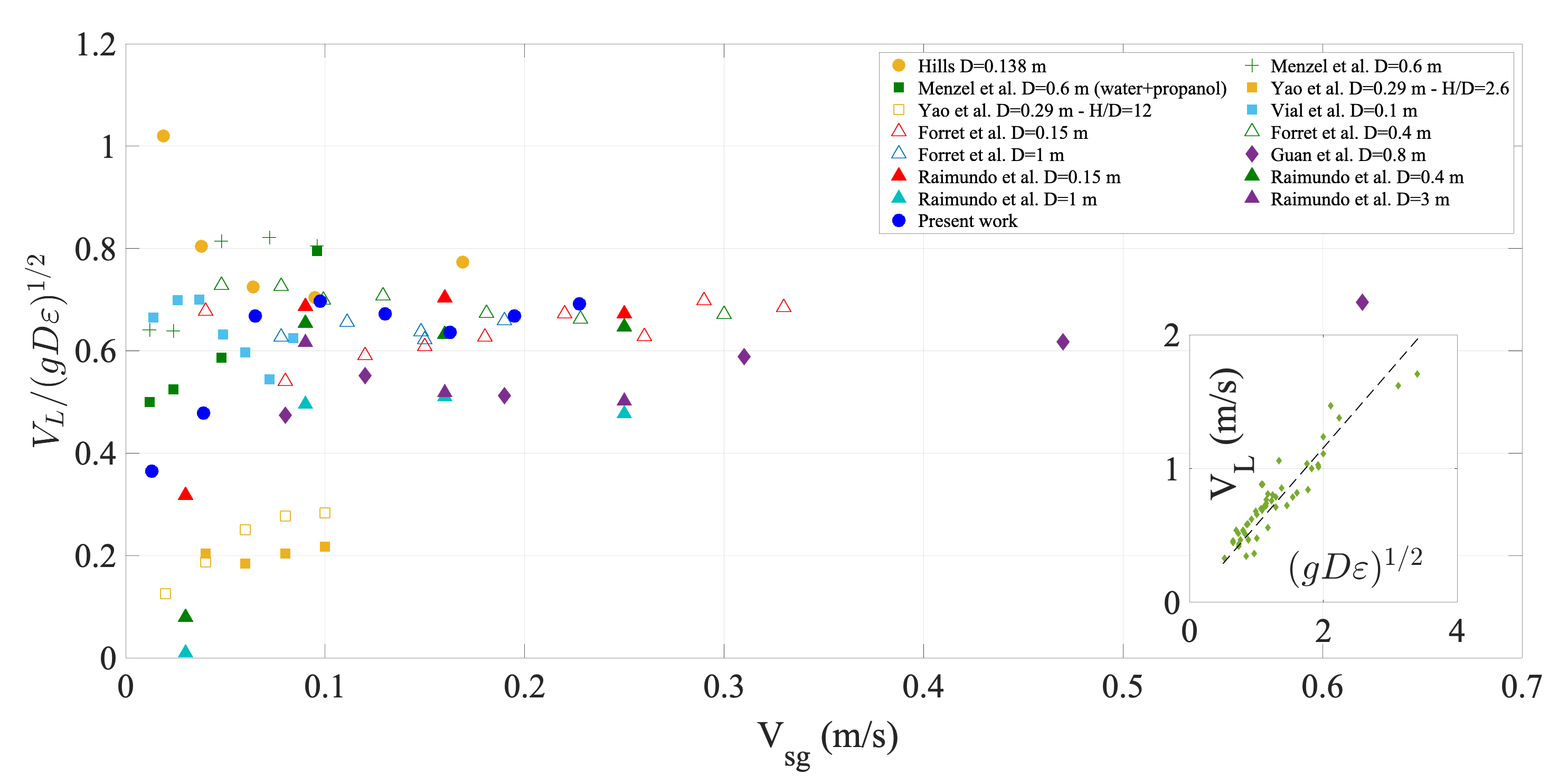}
\caption{Evolution of $V_L/(gD\varepsilon)^{1/2}$ where $V_L$ and $\varepsilon$ are measured on the column axis versus the superficial gas velocity from the contributions quoted in tables \ref{tab1a}\&\ref{tab1b}. The insert plots $V_L$ versus $(gD\varepsilon)^{1/2}$ for all data collected on the column axis in the heterogeneous regime for $V_{sg} \geq 8$cm/s and for $0.1m \leq D \leq 3m$: the dash line in the insert corresponds to the fit $V_L= 0.577 (gD\varepsilon)^{1/2}$.} \label{fig8}
\end{figure}

Figure \ref{fig8} provides the quantity $V_L/(gD\varepsilon)^{1/2}$ versus the superficial gas velocity with $V_L$ and $\varepsilon$ measured on the axis. The figure includes all the contributions listed in tables \ref{tab1a}\&\ref{tab1b}, and that have been collected in the homogeneous regime as well as in the heterogeneous regime. The data concern column diameters from 0.1 to 3m and superficial velocities varying from 1.2 to 62cm/s. In the limit of small superficial gas velocities, $V_L/(gD\varepsilon)^{1/2} $evolves between 0.01 and 1.02. As $V_{sg}$ increases, the range of variation of $V_L/(gD\varepsilon)^{1/2}$ smoothly diminishes. Above $V_{sg} \sim 15$cm/s, $V_L/(gD\varepsilon)^{1/2}$ evolves within the interval $[0.48 ; 0.77]$. Note that this interval corresponds to bubble column diameters ranging from 0.1m to 3m. Above $V_{sg} \sim 20$cm/s, that interval further narrows and the quantity $V_L/(gD\varepsilon)^{1/2}$ tends to be constant in the pure heterogeneous regime: that feature supports the scaling argument presented in Section \ref{sec2}. Although experimental data are lacking at very large $V_{sg}$ to precisely define the asymptotic behavior, the latter is estimated as: 

\begin{equation}\label{eq7}
V_L\sim0.58(gD\varepsilon)^{1/2},
\end{equation}			
as shown by the insert in Figure \ref{fig8} that provides $V_L$ versus $(gD\varepsilon)^{1/2}$ for all available data at $V_{sg}$ above 8cm/s. The proportionality factor equals 0.5774 (and the correlation coefficient is 0.867). Note that the same plot using all data available at $V_{sg}$ above 6cm/s also provides a linear behavior with a proportionality factor equals to 0.5786  (and a correlation coefficient of 0.87). Alternately, when the quantity $V_L/(gD)$ is plotted versus the local void fraction for all the data shown in Fig. \ref{fig8}, $V_L/(gD)$ is found to evolve as $\varepsilon^{0.50\pm0.01}$: this is fully consistent with the $\varepsilon^{1/2}$ dependency predicted.  

\begin{figure}
\centering
\includegraphics[width=\textwidth]{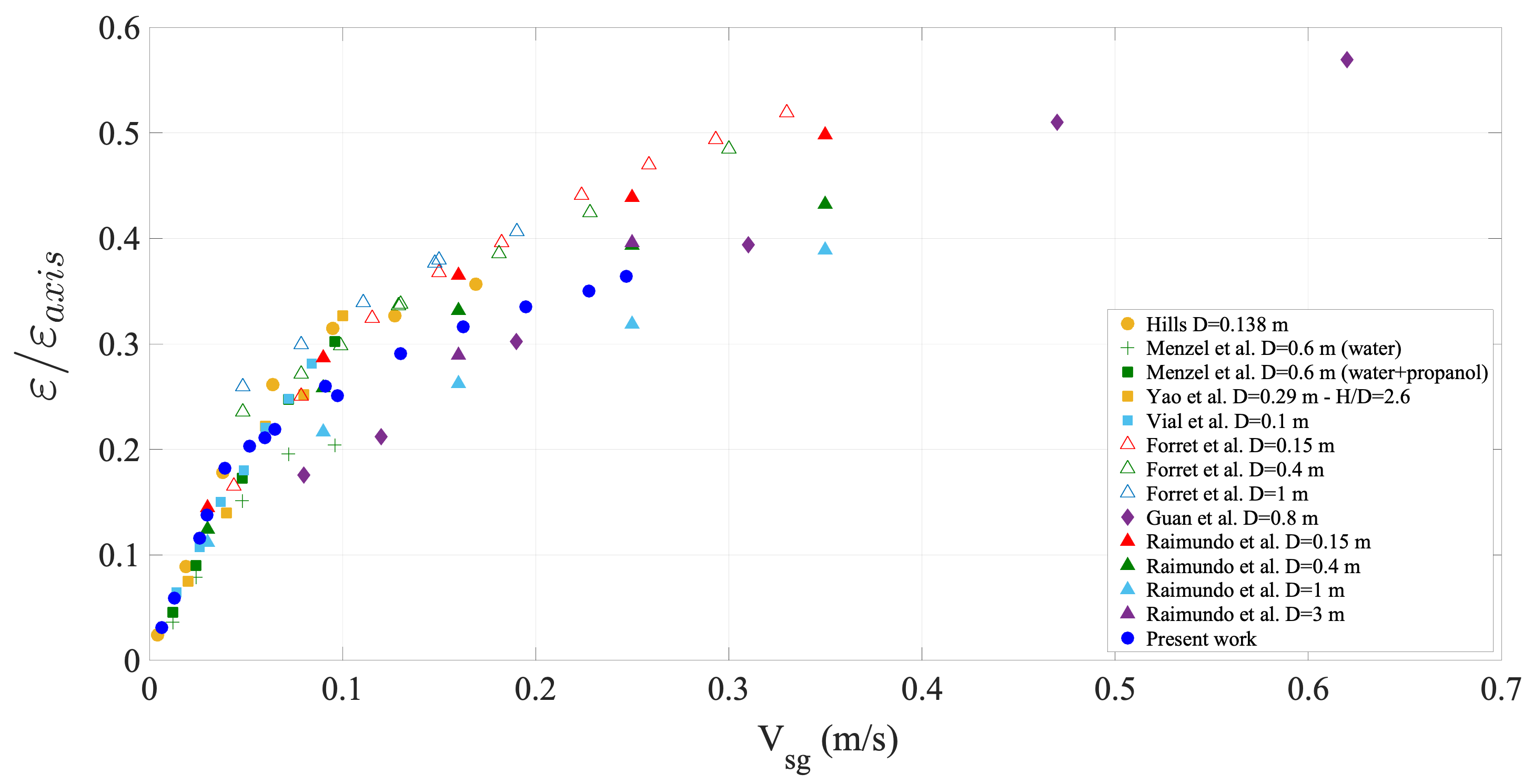}
\caption{Evolution of the local void fraction on the column axis versus the superficial gas velocity for all the contributions quoted in tables \ref{tab1a}\&\ref{tab1b} and exploited in Figure \ref{fig8}. For \cite{forret2006scale}, the local void fraction has been estimated as the global void fraction divided by 1.5.} \label{fig9}
\end{figure}

It is worth discussing the uncertainty on these figures as the data presented in Figure \ref{fig8} come from different operators and from various measuring techniques. The typical dispersion in the measurements can be appreciated by comparing the data collected in identical bubble columns. For example, in the $D=0.15$m column and at $V_{sg} \sim 20cm/s$, there is a 0.2 difference in $V_L/(gD\varepsilon)^{1/2}$ between the data from \cite{forret2006scale} and those from \cite{raimundo2019hydrodynamics}. Similarly, in the $D=0.4$m column and at $V_{sg}\sim20$cm/s, the values of $V_L/(gD\varepsilon)^{1/2}$ deduced from the data of \cite{forret2006scale}, from those of \cite{raimundo2019hydrodynamics} and from the present work all fall within a 0.1 band. These are quite reasonable dispersions especially when considering that the water quality was not always the same, so that the coalescence efficiency varied. Finally, as far as we can judge from the available information given in articles, all the experimental conditions in Figure \ref{fig8} correspond to no or weak coalescence. The monotonous allure of the evolution of the local void fraction on the axis with $V_{sg}$ shown in Figure \ref{fig9} also supports that statement.

\subsection{Mean gas velocity on the column axis \label{sec42}}

Experiments providing statistics on the bubble velocity are not common. This is due to the lack of reliable measuring techniques giving access to bubble velocity in the difficult conditions encountered in the heterogeneous regime, in particular with respect to high void fractions, flow unsteadiness and `chaotic' 3D trajectories of bubbles. Each bubble velocity technique has its own limitations, and their respective uncertainty and resolution in such flow conditions are not well known. For example, for phase detection techniques based on immersed probes either single, double or multiple, it is well known that erroneous velocity data are collected in heterogeneous conditions because of the unsteady 3D motions of these two-phase flows. Yet, average quantities relative to velocity or size seem to be meaningful (\cite{chaumat2007reliability,raimundo2016new}). Moreover, while some bias may alter the statistics and when the latter occurs, it is usually hard to quantify. For all these reasons, we will not discuss further the respective merits of each measuring technique: instead we report all available raw data as given in the original papers, keeping in mind that some issues on resolution, accuracy or bias remain as an open question. 

\begin{sidewaystable}%
  \vspace{18cm}
    \caption{\NOTEN{List of references and flow conditions exploited to extract the gas velocity and local void fraction measured on the column axis. Further information is provided in table \ref{tab2b}}.}
  \vspace{-1.5cm}
  \label{tab2a}
  \includegraphics[width=1.1\textwidth]{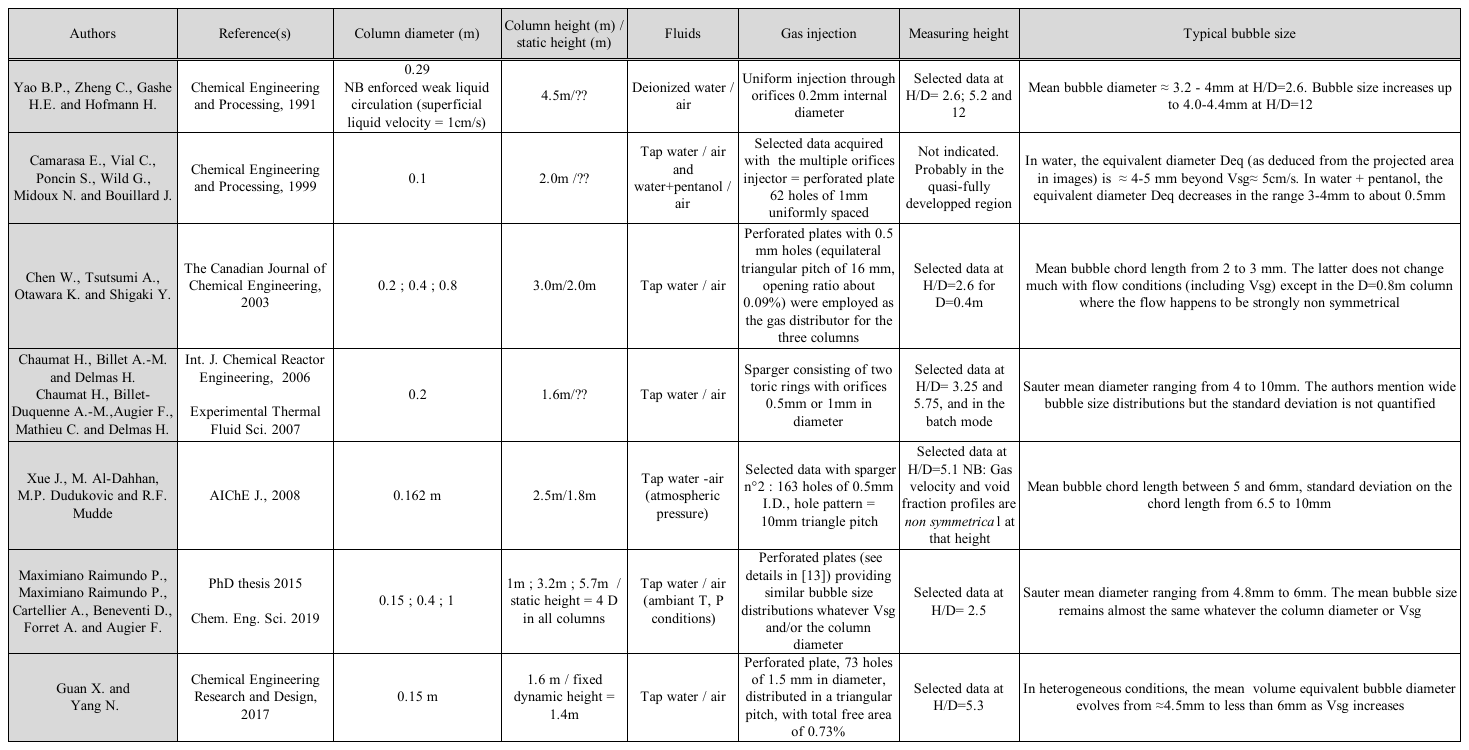}
\end{sidewaystable}

\begin{sidewaystable}%
  \vspace{18cm}
    \caption{\NOTEN{List of references and flow conditions exploited to extract the gas velocity and local void fraction measured on the column axis (this table complements the information given in table \ref{tab2a}).}}
  \vspace{-1.5cm}
  \includegraphics[width=1.1\textwidth]{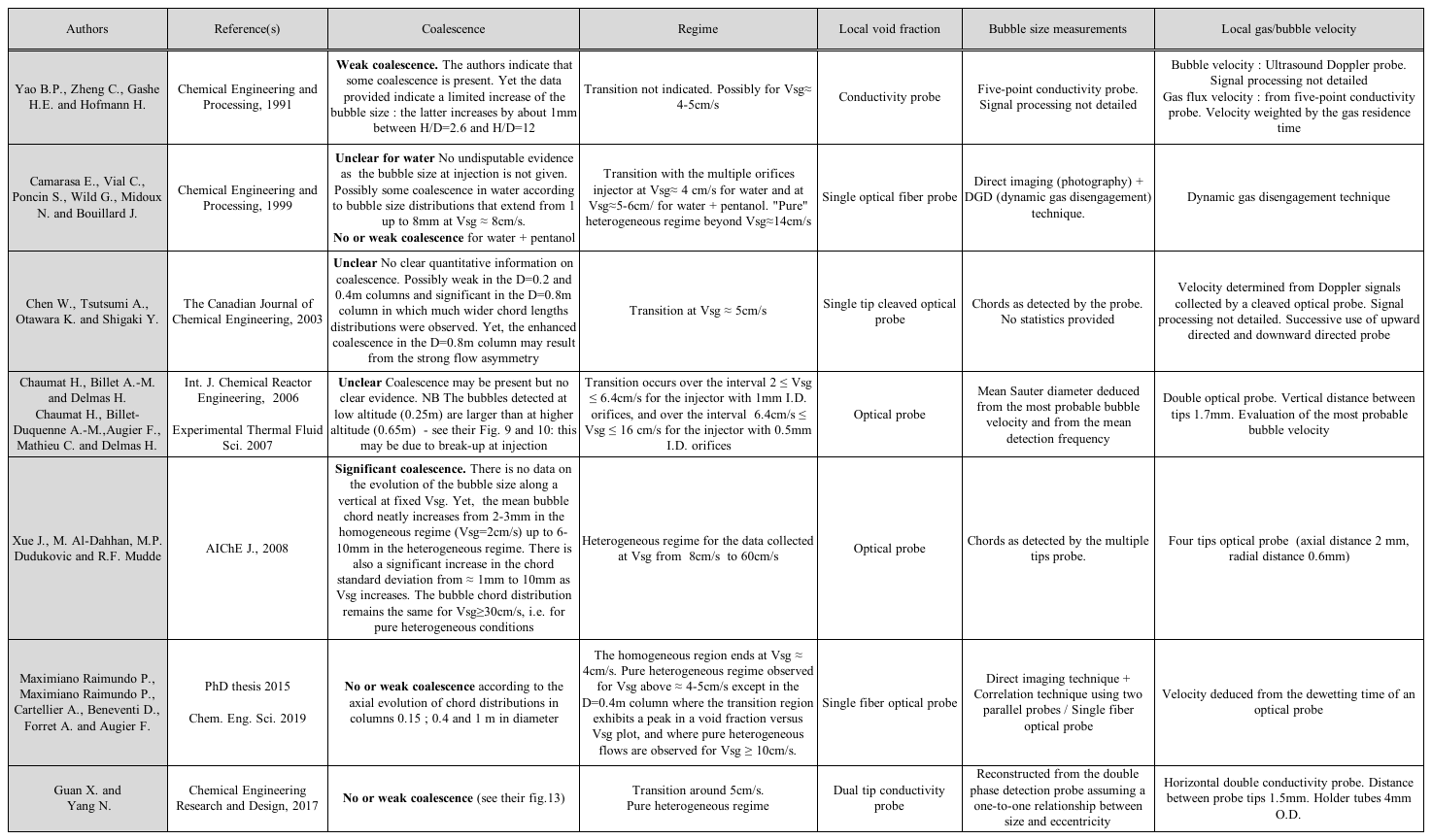}
\label{tab2b}
\end{sidewaystable}

\begin{figure}
\centering
\includegraphics[width=\textwidth]{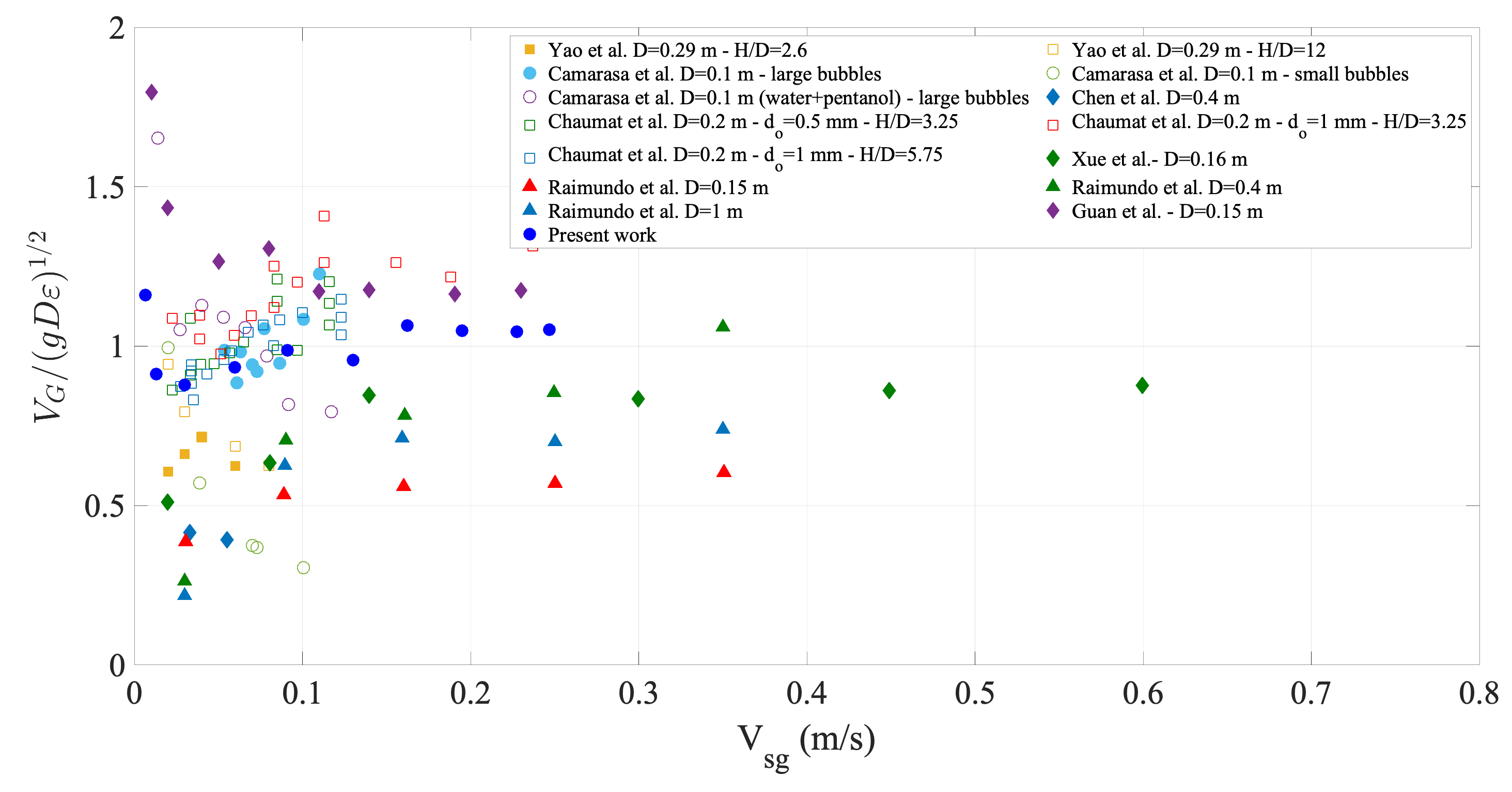}
\caption{Evolution of $V_G/(gD\varepsilon)^{1/2}$ where $V_L$ and $\varepsilon$ are measured on the column axis versus the superficial gas velocity $V_{sg}$ for all gas or bubble velocity measurements from the articles quoted in tables \ref{tab2a}\&\ref{tab2b}.} \label{fig10}
\end{figure}

Tables \ref{tab2a} \& \ref{tab2b} summarize the set of experiments used to analyze the gas velocity that corresponds here to the velocity of bubbles. The corresponding data on the mean bubble velocity scaled by $(gD\varepsilon)^{1/2}$, where both the velocity and the void fraction were measured on the column axis, are reported versus $V_{sg}$ in Figure \ref{fig10} irrespective of the flow regime. 

\NOTEN{ We will now provide some information on how the data were exploited for the datasets used in this section}. First, when different injectors were tested, we always selected the data acquired with multiple-orifice injectors distributed over the entire column cross section. Second, original data were sometimes interpolated to estimate missing local void fraction data: that process was used only when the interpolation process was safe. If some extrapolation was required, the corresponding data were discarded unless otherwise specifically stated in the text in the legend. It is worth to notice that almost all data correspond to tap water and air (under ambient pressure and temperature conditions), with two exceptions: \cite{yao1991bubble} used deionized water, and one set of data from \cite{camarasa1999influence} was gathered in an aqueous solution of alcohol (water and pentanol at a concentration $4  \times 10^{-4}$ mol/l). 

We also mention some specific choices we made when extracting the data. For \cite{yao1991bubble}, only the bubble velocity detected by the ultrasound technique is plotted because these authors found comparable results with their five-point conductivity probe. In \cite{camarasa1999influence}, the bubble velocity was measured by an ultrasonic Doppler technique for single orifice and porous plate gas injection but not for the multiple orifice nozzle that provides the injection conditions we are looking for here. For that sparger, the bubble velocity was measured by a DGD (dynamic gas disengagement) technique: \cite{camarasa1999influence} report the velocity of `large' bubbles and of `small bubbles' but they do not explicitly specify how `large' bubbles are distinguished from `small' ones. It happens that, for a given flow condition, the mean velocity of `small bubbles' measured by \cite{camarasa1999influence} is 2 to 4 times (in water) and 2 to 2.5 times (in aqueous solution of alcohol) lower than the mean velocity measured for `large bubbles'. Presumably, `small' bubbles correspond to the 1-2mm bubbles at the lower end of the bubble size distributions they provide, while `large' bubble can be as large as 8-10mm in water (see their Fig.10) and 6-8mm in water plus pentanol (see their Fig.20). According to these comments, only the data for `large' bubbles are reported in Figure \ref{fig10}. 

Let us finally mention that, for the data of \cite{camarasa1999influence} gathered in the aqueous solution of pentanol, no data is given on the local void fraction and we used the global void fraction instead. Accordingly, the values of $V_G/(gD\varepsilon)^{1/2}$ are overestimated for that series. \cite{chen2003local} performed measurements in a $D=0.4$m bubble column at $H/D=2.6$, and in a $D=0.8$m bubble column at $H/D=1.3$: the later case, for which they observed non-symmetrical flows, was discarded because the data were not collected in the quasi-fully developed region. 

Concerning the experiments by \cite{xue2008bubble}, all the data they collected in the heterogeneous regime at $H/D=5.1$ correspond to strongly asymmetrical flows. Indeed, the velocity difference between the upward motion on one side of the column and the downward motion on the opposite side ranges between 20cm/s and 40cm/s: these figures are therefore quite significant compared with the mean bubble motion on the column axis that evolved between 40cm/s and 90cm/s. That asymmetry was further confirmed by void fraction and bubble detection frequency profiles. The only symmetrical bubble velocity profile reported by \cite{xue2008bubble} was collected at a larger distance from injection (namely $H/D=8.5$) and at $V_{sg}=30$cm/s. Unfortunately they do not provide the void fraction for these conditions. Despite these shortcomings, all the data of \cite{xue2008bubble} collected at $H/D=5.1$ have been integrated in our analysis. Let us also underline that these authors are among the very few who have explored large gas superficial velocities. 

Among the contributions listed in tables \ref{tab2a}\&\ref{tab2b}, almost all gathered (at least in principle) positive and negative bubble velocities, except for \cite{raimundo2015analyse,raimundo2016new} who collected positive velocities only since they exploited the dewetting of a single fiber tip. Yet, as the sources of bias are usually not analyzed for bubble velocity measurement techniques, it is difficult to ascertain that the information collected was indeed the faithful assembly of positive and negative realizations. An indication of these difficulties is that the standard deviation of bubble size distributions are never provided, nor discussed, except by \cite{yao1991bubble} who measured bubble velocity fluctuations with an ultrasound technique. As for the liquid phase, and to be consistent, we thus consider our data series from Section \ref{sec3} named `up and down flow' for the comparison.

The data presented Figure \ref{fig10} concern column diameters from 0.1 to 1m and gas superficial velocities varying over two decades from 0.6 to 60cm/s. At low gas superficial velocities, say below a few cm/s, that is within the homogeneous regime, the quantity $V_G/(gD\varepsilon)^{1/2}$ evolves from 0.2 up to 1.8. That ratio significantly varies from one experiment to the other. For a given data series, the ratio $V_G/(gD\varepsilon)^{1/2}$ tends to become somewhat constant when moving towards large superficial velocities. For $V_{sg}$ above around 10cm/s, that is well within the heterogeneous regime, the dispersion of the data significantly diminishes and $V_G/(gD\varepsilon)^{1/2}$ evolves inside a narrower band comprised between 0.6 and 1.4. Note that these last figures encompass bubble column diameters ranging from 0.15 to 1m. The trends are therefore the same as those detected when analyzing the liquid velocity. Yet, the fluctuations observed from one series to another are larger than those observed for the liquid velocity. This is probably because of the stronger uncertainties of gas velocity measuring techniques. Also, and compared with the mean liquid velocity presented Figure \ref{fig8}, it is more difficult to estimate the asymptotic value of $V_G/(gD\varepsilon)^{1/2}$. According to \cite{xue2008bubble} and to some runs of \cite{raimundo2019hydrodynamics}, that limit is near 0.8 while from the present data, as well as from those of \cite{yao1991bubble}, the limit is possibly closer to 1. Again, experimental data at very large $V_{sg}$ are required to more accurately determine the asymptotic behavior of $V_G/(gD\varepsilon)^{1/2}$. Despite the limitations on available data, the trends observed on $V_G/(gD\varepsilon)^{1/2}$ for column diameters between 0.1 and 1m are consistent with the scaling argument proposed in Section \ref{sec2}.

\begin{figure}
\centering
\includegraphics[width=\textwidth]{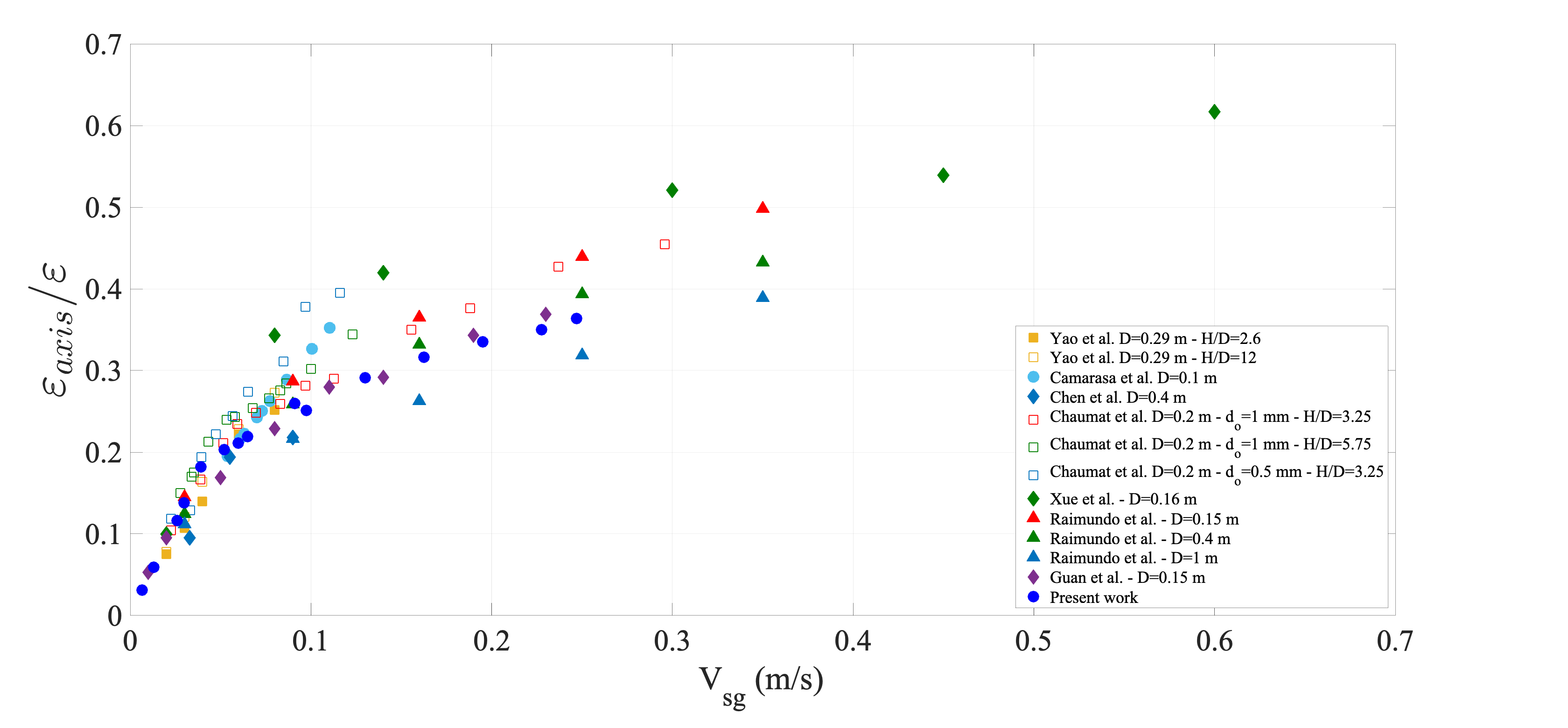}
\caption{Evolution of the local void fraction on the column axis versus the superficial gas velocity for contributions quoted in tables \ref{tab2a}\&\ref{tab2b} and exploited in Figure \ref{fig10}.} \label{fig11}
\end{figure}

Figure \ref{fig11} provides the evolutions of the void fraction on the axis with $V_{sg}$ for all the experiments quoted in tables \ref{tab2a}\&\ref{tab2b}. As in Figure \ref{fig9}, all these evolutions are monotonous, so that they presumably all correspond to no or weak coalescence. Yet, in terms of local void fraction, one data series happens to be neatly above all others: this is the one from \cite{xue2008bubble}. Contrary to all others experiments presented in Figure \ref{fig10} (see also tables tables \ref{tab2a}\&\ref{tab2b}), the coalescence was probably significant in their experiments. Indeed, they observe an increase of the mean bubble chord length on the axis from 2-3mm in the homogeneous regime up to 6mm in the heterogeneous regime while the standard deviation of the size distribution growths from about 1mm up to 10mm. Also, their bubble chord distributions indicate that bubbles up to 15mm are detected, and bubbles on the axis are significantly larger than near walls where their mean chord is less than 3mm. Let us also underline that the local void fractions measured by \cite{xue2008bubble} in the heterogeneous regime are among the largest of all data series presented Figures \ref{fig9} and \ref{fig11}. \cite{xue2008bubble} used injectors made of orifices 0.5mm in diameter, and small orifices are known to produce smaller bubbles and thus to increase the maximum void fraction reached in the homogeneous regime (\cite{joshi1998gas}). 

\cite{chaumat2006axial} obtained similar large values of the local hold-up when using the same small orifices: yet, the values of $V_G/(gD\varepsilon)^{1/2}$ deduced from their measurements remain comparable with others contributions for $V_{sg}$ below 0.13m/s (Fig. \ref{fig10}). These comments indicate that a characterization of flow conditions with respect to coalescence is not easy. In addition, experimental data on bubble velocity are missing to evaluate how much the magnitude of $V_G/(gD\varepsilon)^{1/2}$ may vary with the coalescence efficiency.

So far, the scaling of the liquid velocity has been discussed based on data collected on the axis. For the liquid, it is known that, in the heterogeneous regime, both the transverse liquid velocity and the local void fraction profiles assume self-similar shapes when scaled by their respective value on the axis (\cite{forret2006scale}). Hence, all the above findings are expected to remain valid at any radial position in the column provided one remains in the quasi-fully developed region. For the gas phase, we have shown in \cite{lefebvre2022new} that the bubble velocity profiles collected in the quasi-fully developed region when in heterogeneous conditions also happen to be self similar when scaled by the bubble velocity on the column axis. Hence, the proposed scaling is also expected to remain valid at any radial position for the gas phase. 

\subsection{Liquid and gas velocity fluctuations \label{sec43}}

Experimental data on velocity fluctuations collected in bubble columns are scarce, and this is particularly true in the heterogeneous regime. For the liquid phase, \cite{menzel1990reynolds} provide two profiles of the axial liquid velocity fluctuations (quantified here by the standard deviation $V'$ of the velocity distribution). Nevertheless, these datasets were gathered in a 80 wt\% glycerol/water mixture, and, unfortunately, the authors do not indicate the corresponding mean velocities and void fractions for that fluid. Otherwise, data for the liquid phase are available in \cite{yao1991bubble,vial2001influence,forret2003,raimundo2019hydrodynamics} and from the present contribution. All these data concern deionized or tap water. 

For the gas phase, single tip or multiple tips probes that exploit a transit time technique for velocity measurements are common, but, with these techniques, the measured velocity distributions are strongly biased by the detection of erroneous, large velocities (see for e.g. \cite{chaumat2007reliability,raimundo2016new}). It happens that the mean velocity is significant, but the standard deviation is not reliable. Also, some authors (such as \cite{chen2003local,xue2008bubble}; \cite{guan2017bubble}) provide velocity distributions but the standard deviations are not quantified. For these reasons, one is left only with the data from \cite{yao1991bubble} acquired from an ultrasound technique in a $D=0.29$m column with de-ionized water / air as fluids, and the data we collected with the Doppler probe in a $D=0.4$m column with tap water /air as fluids (see \cite{lefebvre2022new}).  

\begin{figure}
\centering
\includegraphics[width=\textwidth]{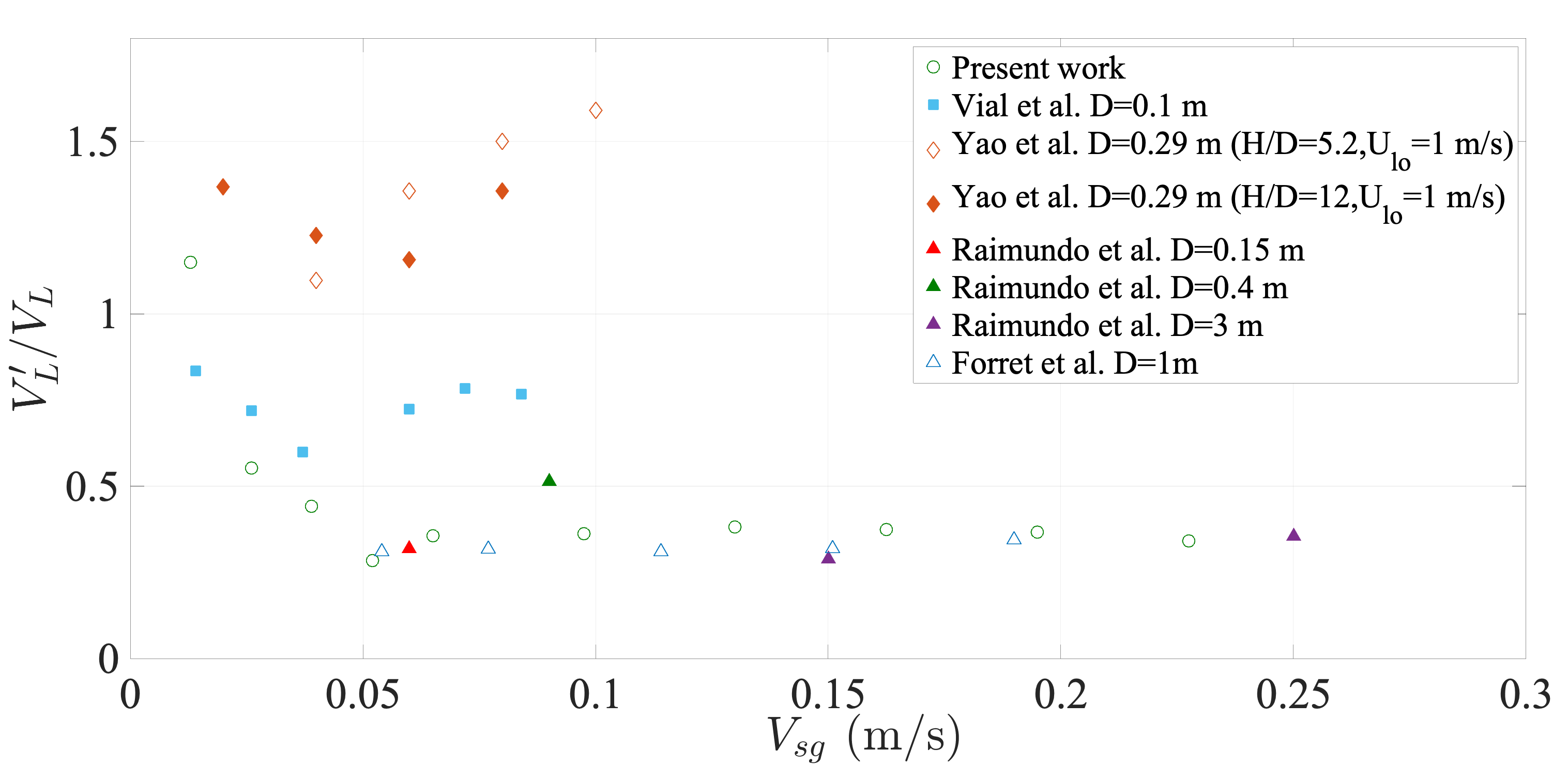}
\caption{Evolution of the relative fluctuation in the liquid phase velocity $V_L'/V_L$ measured on the column axis versus the superficial gas velocity.} \label{fig12}
\end{figure}

\begin{figure}
\centering
\includegraphics[width=\textwidth]{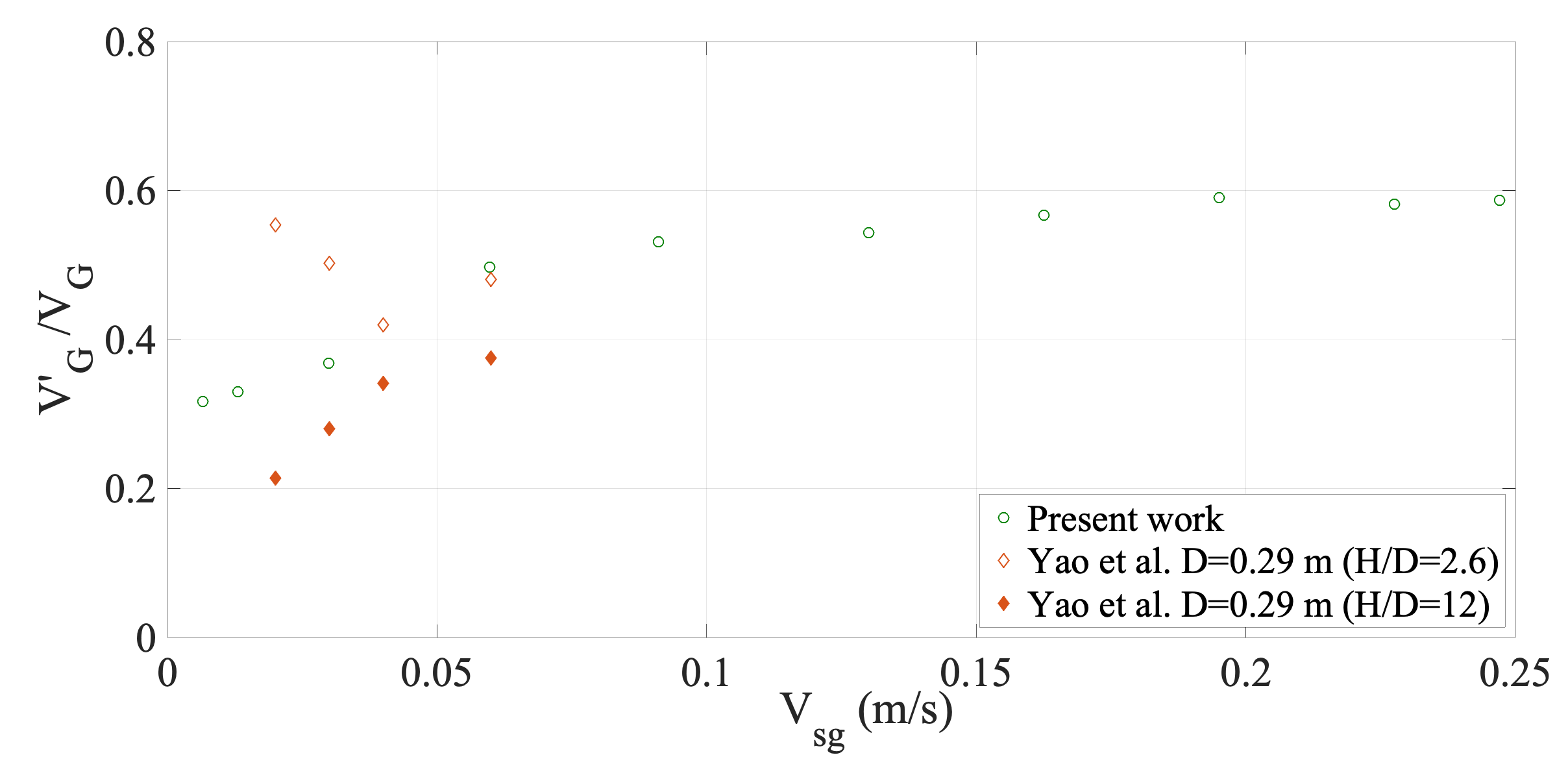}
\caption{Evolution of the relative fluctuation $V_G'/V_G$ of the velocity of bubbles measured on the column axis versus the superficial gas velocity.} \label{fig13}
\end{figure}

The experiments of \cite{menzel1990reynolds} indicate that, in the heterogeneous regime, the radial profiles of liquid velocity fluctuations remain self-similar when normalized by their maximum. One can also use the velocity fluctuations evaluated on the axis of the bubble column for that normalization. Hence, in the following, we focus our discussion on velocity fluctuations measured on the axis. The relative fluctuation $V_L'/V_L$ in the liquid phase (respectively $V_G'/V_G$ for the gas phase) measured on the column axis is presented Figure \ref{fig12} (respectively Figure \ref{fig13}) versus the superficial gas velocity. All these measurements have been done in the quasi-fully developed region. Despite the limited number of independent data, each of the quantities $V_L'/V_L$ and $V_G'/V_G$ tends towards a constant value when moving inside the heterogeneous regime. That feature is well established for the liquid phase since the data come from different operators and from different sensors. Remarkably, the asymptotic behavior is the same whatever the column diameter ranging from 0.15 to 3m. 
The only series exhibiting a different trend are those from \cite{yao1991bubble} for the liquid phase. These data were obtained from hot-film probes, a technique that could be delicate to exploit in bubbly flows. The difficulties are expected to be even stronger in the conditions encountered in bubble columns at high gas superficial velocities. Unfortunately, \cite{yao1991bubble} do not comment on the signals they collected, nor on the signal processing they develop: it is therefore difficult to evaluate the reliability of their measurements.

Figures \ref{fig12} and \ref{fig13} show that, in the heterogeneous regime, the velocity fluctuations in the liquid as well as in the gas remain proportional to the mean velocity. Hence, all the findings on the scaling of mean velocities in the core region of the bubble column also apply to velocity fluctuations. One may be puzzled by such result, but physical arguments similar to those evoked in section \ref{sec2} can explain that feature. The idea is that, in the heterogeneous regime, the contributions to velocity fluctuations in the liquid phase by the relative motion at the bubble scale, i.e. the so-called bubble induced-turbulence, which is at the origin of large ratio $V_L'/V_L$ observed Fig. \ref{fig12} in the homogeneous regime (see \cite{risso2018agitation}), is not the leading mechanism. Instead, in the heterogeneous regime, the agitation in the liquid is due to the presence of meso-scale structures. The later have been put in evidence and quantified with a 1D Vorono\"i analysis performed on the phase indicator function delivered by optical probes. With such Vorono\"i tesselations, we have shown that in the heterogeneous regime these flows as organized in clusters (high void fraction regions) and voids (low void fraction regions) (\cite{raimundo2019hydrodynamics}). The variations in void fraction from one meso-scale structure to the other induce buoyancy forces that spatially fluctuate, and hence a velocity field that changes from one structure to the other. These meso-scale structures are transported by the mean flow and they could also form and disappear. Hence, at a fixed point in space, the passage of successive structures induce the velocity fluctuations that are precisely those detected by an Eulerian measuring technique, i.e. they are the quantities $V_L'$ and $V_G'$ measured with local probes. 

\section{Velocity scaling: further considerations \NOTE{on the void fraction prediction} \label{sec6}}

We have shown that, in the heterogeneous regime, all velocities, namely the mean velocity and its standard deviation in both phases as well as relative velocity between phases, scale with $(gD\varepsilon)^{1/2} $ as expected from an inertia-buoyancy equilibrium. We have shown that this scaling holds in a $D=0.4$m column for all the flow conditions pertaining to the heterogeneous regime that we have investigated. The analysis of literature data has confirmed the validity of the proposed scaling for the mean and for the standard deviation of the liquid velocity in columns of diameter comprised between 0.1m and 3m, for the mean gas velocity in columns of diameter between 0.1m and 1m, and for the standard deviation of the gas velocity in columns of diameter $D=0.29$m and 0.4m. For the relative velocity, the only data available concern the $D=0.4$m bubble column considered here. Let us recall that almost all flow conditions analysed correspond to air-water systems with large bubbles at high particle Reynolds numbers (Section \ref{sec4}).

To reach a fully predictive status for velocities, one also needs the gas hold-up $\varepsilon$ as a function of flow parameters. However, there is no consensus on the void fraction prediction in bubble columns operated in the heterogeneous regime as tens of different correlations involving various sets of parameters are proposed in the literature (as notably shown by the reviews by \cite{joshi1998gas,kantarci2005bubble,kikukawa_physical_2017,besagni_two-phase_2018}). There is even no clear consensus on the set of non-dimensional parameters governing the response of the system. We propose in Appendix \ref{appA} a dimensional analysis dedicated to the heterogeneous regime that is restricted to high aspect ratio bubble columns, to systems far from critical conditions, and without or with weak coalescence. We identify five independent non-dimensional parameters, and a possible choice could be:

\begin{itemize}

\item	the Archimedes number $Ar=gD^3  / \nu_L^2$, $Ar$ is the square of a Reynolds number based on the column diameter $D$, on the liquid viscosity and on the velocity scale $(gD)^{1/2}$.

\item	the Froude number $Fr = V_{sg}/(gD)^{1/2}$ , that quantifies the injected gas flow rate.

\item	E\"otv\"os number $Eo=\rho_L g d^2/\sigma$, that measures the mean bubble size d relative to the capillary length. 

\item	Morton number $M = g \mu_L^4 / (\rho_L \sigma^3)$, that involves the physical properties of the couple of fluids selected and the gravitational acceleration. 

\item	a non-dimensional parameter quantifying the degree of polydispersity in the system defined as the standard deviation of the bubble size distribution $std(d)$ divided by the mean bubble diameter $d$.

\end{itemize}

Among these, the E\"otv\"os and Morton numbers completely define the dynamics, that is the shape, the trajectory and the relative velocity, of an isolated bubble having the mean equivalent diameter $d$ immersed in the stagnant liquid and for the given gravitation field (strictly speaking, this only holds for clean interfaces). 

All the experimental conditions analysed here (see section \ref{sec4}) involve the same couple of fluids (i.e. air and water in ambient thermodynamic conditions) and earth gravity so the $M$ parameter is the same ($\sim 10^{-11}$). The $Eo$ parameter evolves in a rather narrow range, roughly from 1 to 10. For these $M$ and $Eo$ values, the bubbles are in the so-called wobbling regime, and they have similar dynamics with $O(10^3)$ particle Reynolds numbers (see Section \ref{sec4}). The polydispersity parameter $std(d)/d$ is scarcely quantified but, according to available bubble size distributions, it does not change much (the minimum bubble size is typically of the order of $0.5-1$mm while the maximum bubble size never exceeds $\sim$10mm). Hence, in the experiments quoted in tables \ref{tab1a}, \ref{tab1b}, \ref{tab2a} and \ref{tab2b}, only the two parameters $Ar$ and $Fr$ have been significantly changed. The $Ar$ number evolves between $9.8 \times10^9$ and $2.6 \times 10^{14}$. 

\begin{figure}
\centering
\includegraphics[width=0.8\textwidth]{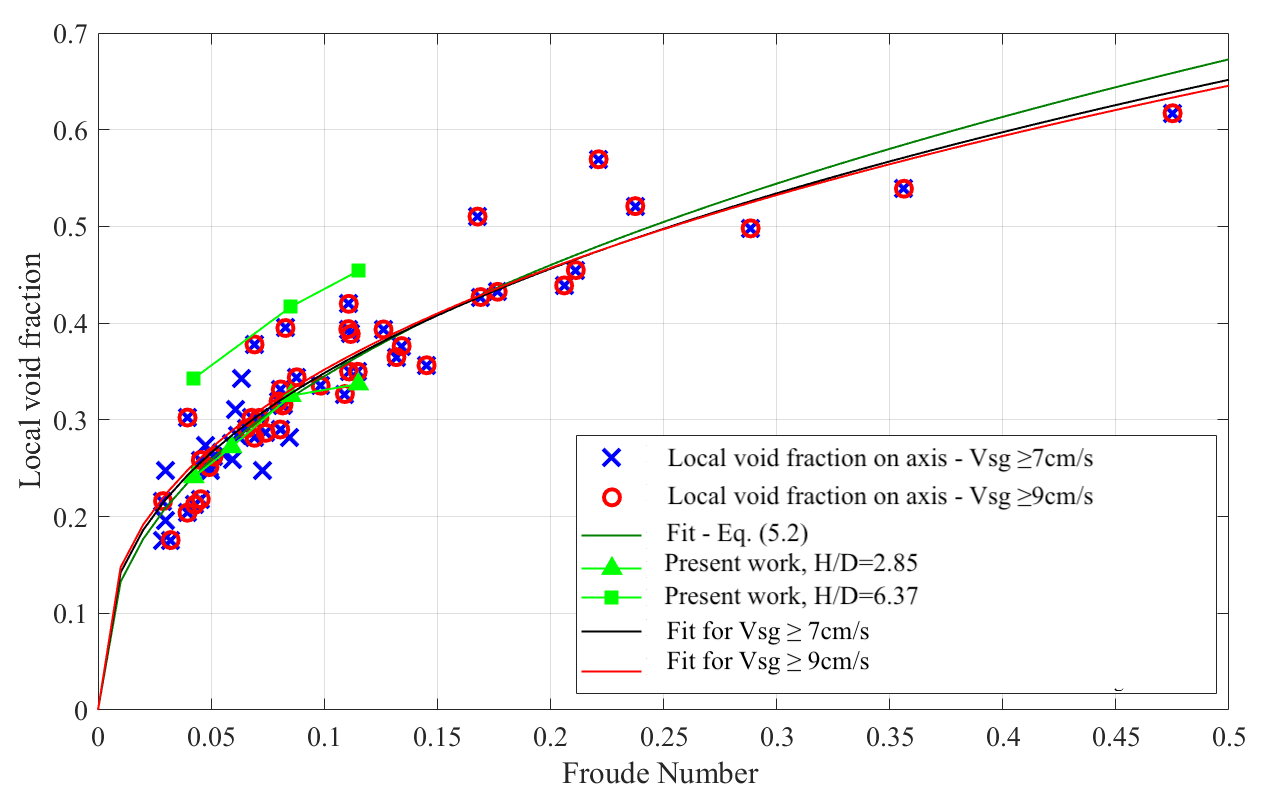}
\caption{Correlation between the local void fraction on the axis of the column and the Froude number Fr=$V_{sg}/(gD)^{1/2}$ for all experiments quoted in tables 1 and 2. Measurements performed for $1.3 \leq H/D \leq 12$. To illustrate the impact of H/D, the data with closed symbols correspond to measurements in the D=0.4m column performed at moderate (H/D=2.85, triangles) and at large (H/D=6.37, squares) distances from injection.} \label{fig30}
\end{figure}

Furthermore, to be sure to analyse data pertaining to the heterogeneous regime (and thus to escape from the transition zone), let us consider gas superficial velocities above 7cm/s or 9cm/s. The corresponding Froude numbers span the range $[0.016; 0.475]$. According to Figures \ref{fig7} and \ref{fig9}, the void fraction seems to be mainly driven by the superficial gas velocity $V_{sg}$. To check that, we first attempted a correlation with both $Fr$ and $Ar$, and an exposant as low as 0.047 was found for the $Ar$ number. We therefore examined how the void fraction evolves with the Froude number alone. As shown Fig. \ref{fig30}, the local void fraction correlates well with $Fr$ as one gets:

\begin{equation}\label{eq17}
\begin{split}
\varepsilon_{axis}  = 0.853 Fr^{0.389}, ~ V_{sg} \geq 7 cm/s, \\
\varepsilon_{axis} = 0.838 Fr^{0.377}, ~ V_{sg} \geq 9 cm/s,
\end{split}
\end{equation} 

\noindent with correlation coefficients of about 0.8. The maximum deviation of these fits from measurements is $\pm$30\% except for two data collected in a $D=3$m column at $V_{sg}=16$cm/s and 25cm/s for which the deviation reaches 35\%. The measurements in large columns are not easy (probably due to vibration of the probe holder, and/or to flow perturbation induced by the latter when it is too large). If these two data points are discarded, the correlation becomes:

\begin{equation}\label{eq18}
\varepsilon_{axis} = 0.897 Fr^{0.415},
\end{equation} 

\noindent with a correlation coefficient of 0.887. Eq. \ref{eq18} holds for $V_{sg} \geq 7 $cm/s as well as for $V_{sg} \geq 9$ cm/s. The deviation remains then within $\pm22$\% for all data. The correlation coefficients as well as the maximum deviations found here appear as acceptable, especially if one accounts for the fact that the data considered were collected at different heights above injection (see tables \ref{tab1a}, \ref{tab1b}, \ref{tab2a} and \ref{tab2b}). Indeed, we have shown in \cite{lefebvre2022new} that the local void fraction on the axis linearly increases with the height in the quasi fully developed region of the flow. The impact of the measuring height on the local void fraction is illustrated in Fig. \ref{fig30} by two datasets (closed symbols) gathered on the axis of the $D=0.4$m column: the upper dataset corresponds to $H/D=6.37$ and the lower data set corresponds to $H/D=2.85$. Clearly, the distance between these series is comparable with the dispersion.

The disappearance of the Archimedes number in the above empirical expressions for the void fraction is not unexpected. Indeed, as discussed in Appendix \ref{appA}, the heterogeneous regime in a bubble column corresponds to a turbulent regime in free thermal convection. Thus, the Archimedes number somewhat controls the transition to that regime (the critical Rayleigh number introduced by \cite{ruzicka2003buoyancy} to identify the homogeneous-heterogeneous transition is proportional to $Ar$). However, once the turbulent regime is installed, buoyancy forcing overwhelms viscous effects, and the precise value of $Ar$ is no longer relevant for setting the dynamical equilibrium: this is why its outcome, i.e. the void fraction, is no longer dependent on $Ar$. 

There is also a debate as wether the void fraction should depend or not on the bubble column diameter $D$. For example, according to \cite{besagni_two-phase_2018}, the correlation proposed by \cite{akita1973gas} should be considered as the state of the art for determining the global gas hold-up. That correlation does not include any dependency of the void fraction on $D$. Yet, among the experiments quoted in tables \ref{tab1a}, \ref{tab1b}, \ref{tab2a} and \ref{tab2b}, huge differences (with factors much larger than one) appear between gas hold-up measurements and predictions using the Akita and Yoshida's correlation. Beside, \cite{ruzicka2001effect} unambiguously demonstrated that an increase in the column diameter advances the transition. Therefore, one expects some dependency of the void fraction on the bubble column diameter. According to the fits proposed here, $\varepsilon_{axis}$ evolves as $\sim V_{sg}^{0.4}$ and as $\sim D^{-0.2}$.

Going back to the scaling for the mean liquid velocity (eq. \ref{eq7}) established in section \ref{sec41}, and using $\varepsilon_{axis}=0.9 Fr ^{0.4}$ as a convenient approximation of eq. \ref{eq17} to eq. \ref{eq18}, we obtain:

\begin{equation}\label{eq19}
V_L / (gD)^{1/2} \sim 0.58 \varepsilon^{1/2} \sim 0.55 Fr^{0.2}.
\end{equation} 

\noindent Eq.(19) leads to $V_L  \sim 1.09 V_{sg}^{0.19} D^{0.3}$ which is close to the empirical fit proposed by \cite{raimundo2019hydrodynamics}, that wrote $V_L \sim 1.35 V_{sg}^{0.16} D^{0.4}$. Hence, we recover a formula similar to that empirical fit by exploiting the inertia-buoyancy argument leading to the $(gD\varepsilon)^{1/2}$ velocity scaling, combined with an empirical relationship between the local void fraction and the Froude number. These findings are therefore consistent with each other. 

Let us compare eq. \ref{eq19} with available results from the literature. The majority of the correlations proposed for the mean liquid velocity are not in a dimensionless form. Among those that are dimensionally consistent, only one proposal uses $(gD)^{1/2}$ as a velocity scale. The latter writes:

\begin{equation}\label{eq20}
V_L / (gD)^{1/2} = 0.737 Fr^{1/3}.
\end{equation} 

\noindent Eq. \ref{eq20} was proposed by Zehner (see \cite{zehner1996modelling}) and by \cite{kawase1986liquid} who quote an earlier publication by Zehner in 1982 that provides the same prediction. Eq. \ref{eq20} was derived using modelling considerations: it is based on a simplified axial momentum equilibrium and on a global energy balance where the dissipation in the column is estimated from mean liquid velocity profiles using a mixing length approach. According to \cite{kawase1986liquid}, Eq. \ref{eq20} is reliable for column diameters from 0.1 to 1m, and for Newtonian fluids with a dynamic viscosity between $10^{-3}$ and $2 \times 10^{-2}$ Pa s: the range of $Fr$ is not indicated. Besides, \cite{kawase1987theoretical} correlated the global gas hold-up $R_{G3}$ for Newtonian fluids as:

 \begin{equation}\label{eq21}
R_{G3} = 1.07 Fr^{2/3}.
\end{equation} 

\noindent Let us first underline that the exponents of $Fr$ appearing in equations \ref{eq20} and \ref{eq21} are compatible with the scaling (see eq. \ref{eqscal}) derived from inertia-buoyancy equilibrium. However, the $1/3$ exponent of $Fr$ in eq. \ref{eq20} is significantly larger than the empirical value 0.2 found here (see eq. \ref{eq19}). Similarly, concerning the dependency of the void fraction on the Froude number, the exponent 2/3 found by \cite{kawase1987theoretical}, is somewhat larger than the empirical value ($\sim 0.38-0.4$) found here (see Fig. \ref{fig30}). Let us finally underline that \cite{kawase1987theoretical} found eq. \ref{eq21} valid for $D$ between 0.1 and 1.07m and for $Fr$ from 0.005 to about 0.05. It happens that these flow conditions mostly correspond to, or/and are very close to the homogeneous regime (see \cite{kawase1987theoretical1}), so that the comparison of their proposal with the results we obtained in a pure heterogeneous regime may not be entirely relevant. 

Further investigations are therefore required to accurately determine how the void fraction evolves with non-dimensional parameters, and in fine to predict absolute as well as relative velocities in bubble columns operated in the heterogeneous regime. The above proposal is believed to be a rather robust first step in that direction. 

At this stage, it is relevant to briefly discuss what would be necessary ingredients to elaborate a predictive model. In Appendix \ref{appB}, we present an elementary model for estimating the void fraction in the heterogeneous regime. Following \cite{zuber1965average}, we assume a 1D mean vertical flow in the central part of the column. That central part is considered as an inner tube fed by the liquid flow rate $Q_{Lup}$ set by eq. \ref{eq6} and by a gas flow rate $Q_{Lup} = (1+c) \pi R^ 2 V_{sg}$. Here, $c$ denotes the fraction of the injected gas flow rate $\pi R^ 2 V_{sg}$ that is recirculated. The value of $c$ has been estimated to vary between 0 and 0.2 (\cite{lefebvre2022new}). The gas flow rate fraction $\beta$ injected in that tube is then $Q_{Gup}/(Q_{Gup}+Q_{Lup})$. Using Zuber \& Findlay's approach, we derive a relationship between the gas concentration $\langle \varepsilon \rangle $ in that tube and $\beta$. The difference between $\langle \varepsilon \rangle $ and $\beta$ is controlled by the ratio of the mean gas velocity to the mean liquid velocity, where mean values correspond here to an average over the tube cross-section. Introducing known velocities and void fraction profiles in such a model, the predicted mean void fraction $\langle \varepsilon \rangle $ happens to be in reasonable agreement (within 20\%) with experiments in the $D=0.4$ m column at the beginning of the heterogeneous regime that is up to $V_{sg} \sim 9$cm/s. This is a quite encouraging result, and it success lies on the knowledge of the circulated liquid flow rate, and on direct measurements of the actual relative velocity between phases. In addition, as $\beta$ is only a function of $Fr$ (see eq. \ref{eqB5} in Annex \ref{appB}), that elementary model brings some support to the correlation $\varepsilon(Fr)$ discussed in this section.

\NOTE{However, the deviation between predictions and measurements happens to monotonously increase with $V_{sg}$: it reaches 80\% at $V_{sg}\sim25$cm/s in the $D=0.4$m column (see Fig. \ref{figB2}). A similar trend arises from data extracted from literature for columns diameters between 0.13m and 3m: a reasonable agreement holds at the beginning of the heterogeneous regime but deviations reach a factor about 2 and above at $V_{sg} \geq 40$cm/s (see Fig. \ref{figB3}). }

\NOTE{Our understanding of the reason for these deviations is the following. First, we stress that the Zuber \& Findlay's approach hold for a truly fully developed flow. Second, although self-similarity holds, a few measurements (\cite{lefebvre2022new}) indicate that, in the heterogeneous regime, the void fraction on the axis of the column significantly increases with the distance to injection. These measurements indicate also that the axial void fraction gradient increases with $V_{sg}$. We therefore suspect that such axial evolutions are at the origin of the deviations of the model from reality at large $V_{sg}$. In other words, the 1D assumption of the model needs to be relaxed so that an extra dependancy of the dynamics on $V_{sg}$ could be identified.  It could well be that the downward directed gas flow rate is not fully recirculated at the bottom but that it continuously feeds the central tube. That would render the gas flow rate fraction but also the mean phasic velocities as well as the void fraction dependent on the height. A consequence of that discussion is that much more attention should be paid in future investigations to characterise and to analyse the axial evolutions of key variables.}


\section{Conclusions \NOTE{and perspectives}}

We revisited the hydrodynamics of bubble columns operated in the heterogeneous regime. Conditions with large enough aspect ratio were selected to ensure the presence of a quasi-fully developed region where transverse profiles of void fraction, liquid and bubble mean velocities remain self-similar. We also focused the analysis on air-water systems in ambient thermodynamic conditions involving bubbles in the wobbling regime with large, $O(10^3)$ particle Reynolds numbers. 

\NOTE{We have shown that the dynamical equilibrium in these gravity driven bubbly flows balances liquid inertia with buoyancy. Contrary to thermal convection in tubes that involve an unstable vertical stratification, the vertical density gradient on the axis is stable in bubble columns. Instead, the transport of bubbles induces a transverse gradient in void fraction, so that the driving force of the main motion in bubble columns arises from the radial density distribution. The resulting scaling for velocities is $V\sim(gD\varepsilon)^{1/2}$, where $D$ is the diameter of the bubble column, $\varepsilon$ the void fraction and $g$ the gravitational acceleration.}

\NOTE{Using new experiments performed in a $D=0.4$ bubble column, as well as data extracted from literature, this scaling proposal has been shown to hold for the large-scale motion of both liquid and gas phases. This proposal happens to be valid over a wide range of flow conditions, namely for $D$ from 0.1 to 3m and for gas superficial velocities $V_{sg}$ up to 60cm/s: the corresponding Froude number $Fr = V_{sg}/(gD)^{1/2}$ spans a range from $\sim 0.02$ to $\sim 0.5$. The same scaling applies to velocity fluctuations, as the latter were found proportional to the mean velocity both in the gas and in the liquid. Moreover, we also confirmed the finding of  \cite{raimundo2019hydrodynamics} that the recirculating liquid flow rate $Q_{Lup}$ in the heterogeneous regime is uniquely set by the column diameter, namelly $Q_{Lup} = 0.098 S_{core} (gD)^{1/2}$, where $S_{core}$ is the cross-section of the upward mean flow region centered on the column axis. That result also supports the fact that $(gD)^{1/2}$ is a natural velocity scale for the vertical transport.}

\NOTE{Direct measurements of unconditional mean phasic velocities in the $D=0.4$m column show that the relative velocity levels off at the homogeneous-heterogeneous transition, increases with the gas superficial velocity, and seems to asymptote to $\sim2.4$ times the terminal velocity of bubbles at large $V_{sg}$. Such an analysis deserves to be pursued in order to connect these findings with the known presence of meso-scale structures in the heterogeneous regime. In particular gas velocity measurements conditioned by the local concentration which are now accessible, will help identifying and quantifying collective effects leading to enhanced relative velocities.}

\NOTE{As a prediction of the gas hold-up is still crucially needed, an empirical proposal has been made for air-water systems involving wobbling bubbles, in which the local void fraction $\varepsilon$ is a function of the Froude number alone. We also attempted a void fraction prediction using a Zuber \& Findlay approach: the latter proves successful just after the homogeneous-heterogeneous transition but it fails at larger gas superficial velocities. That feature indicates that the evolutions of key variables such as void fraction, phasics velocities along a vertical are significant even in the quasi-fully developed region of bubble columns. Further efforts should therefore be dedicated to characterise the axial evolutions of variables, and to understand the origin of the global self-organisation prevailing in the heterogeneous regime. }

Finally, a question of importance concerns the impact of coalescence on the above findings. We anticipate that the flow dynamics discussed here and the proposed scalings would remain valid as far as bubbles do not become too large. This statement is already supported by the experiments we analysed as the latter cover different situations in terms of coalescence efficiency. \NOTE{Another argument is the following: the size of a bubble needed for its terminal velocity to equal the asympotic relative velocity measured in the $D=0.4$m column (about 0.7m/s) would be 5cm. Hence, up to that size limit for bubbles, we do not expect the flow dynamics to drastically change because of coalescence. }


\subsection*{Acknowledgements}
The LEGI is part of the LabEx Tec21 (Investissements d'Avenir  - grant agreement n. ANR-11-LABX-0030). That research was also partially funded by IDEX UGA (n. ANR-15-IDEX-0002) The authors report no conflict of interest.


\appendix

\section{Dimensional analysis }\label{appA}

A tentative dimensional analysis of bubble column hydrodynamics may be the following. First, we consider that a quasi-fully developed region does exist when in the heterogeneous regime. We restrict the analysis to situations where coalescence is not playing a key role. More precisely, there is no or weak coalescence in the quasi fully developed region of the column. Coalescence may be present in the entrance region just above injection and thus control the `equilibrium' bubble size distribution, but it is not active in other regions of the column. In such circumstances, the flow in the quasi-fully developed region becomes insensitive to the detail of the injector design (provided some precautions on the design of that device). The relevant physical quantities are the following:

\begin{itemize}

\item $D$ column diameter,

\item $Ho$ static liquid height in the column,

\item $Q_G$ injected volumetric gas flow rate or superficial velocity $V_{sg}= Q_G / (\pi D^2/4)$

\item $g$ gravitation acceleration,

\item $d$ mean bubble size,

\item $std(d)$ standard-deviation on bubble size distribution

\item $\rho_L$ liquid density

\item $\rho_G$ gas density

\item $\mu_L$ liquid dynamic viscosity

\item $\mu_G$ gas dynamic viscosity

\item $\sigma$ surface tension

\end{itemize}

Note that because coalescence is discarded (assuming it has not a significant impact on the flow dynamics when it is weak enough), we do not account for physical quantities such as surface tension gradients nor surfactant concentration and transport and their consequences on interfacial rheology that can affect coalescence efficiency. We keep the standard-deviation $std(d)$ of the bubble size distribution as a parameter. Indeed, some suggestions by \cite{lucas2005influence} supported by experiments from \cite{lucas2019influence} tend to indicate that the extent bubble size distribution has an impact on the transition from the homogeneous to the heterogeneous regime. Also, the approach developed by \cite{krishna1991model} based on a bi-modal bubble size distribution requires $std(d)$ as a parameter. Yet, it is not ascertained that the extent bubble size distribution has an impact on the dynamics in the heterogeneous regime. Without clear evidence in one direction or in the other, that parameter is kept in the list. 

We restrict the analysis to ambient pressure and temperature that is far from critical conditions. Hence, the gas to liquid density and dynamic viscosity ratio remain much smaller than unity: we assume that they have an asymptotic behavior and therefore the two parameters $\rho_G$ and $\mu_G$ disappear from the analysis. 

We are left with 9 physical parameters: $D$, $Ho$, $Q_G$, $g$, $d$, $std(d)$, $\rho_L$, $\mu_L$ and $\sigma$. That list leads to 6 non-dimensional parameters. A possible choice could be:

\begin{itemize}

\item	Archimedes number $Ar=gD^3\rho_L \delta \rho / \mu_L^2=[gD^3/\nu_L^2](\delta \rho/\rho_L)$, with $\delta \rho = \rho_L-\rho_G$. Far from critical conditions, $\delta \rho / \rho_L \sim 1$ and thus, $Ar = gD^3/\nu_L^2$.

\item	Aspect ratio $Ho/D$

\item	Froude number $Fr = V_{sg}/(gD)^{1/2}$

\item	E\"otv\"os number $Eo=\rho_L g d^2/\sigma= (d/a_c)^2$, where $a_c$ is the capillary length scale. 

\item	Morton number $M = g \mu_L^4 / (\rho_L \sigma^3)$

\item	Non-dimensional width of the size distribution $std(d)/d$

\end{itemize}

We have seen that the response of the system does not depend on the static liquid height $Ho$ when the aspect ratio $Ho/D$ is large enough. We are thus left with 5 independent non dimensional parameters, namely $Ar$, $Fr$, $Eo$, $M$ and $std(d)/d$. 

Within that list, the E\"otv\"os and the Morton numbers control the dynamics of an isolated bubble in a quiescent fluid (\cite{clift2005bubbles}): in particular, they control the shape of the bubble and its terminal velocity $U_T$ in the selected fluids and gravity field. Almost all experimental data mentioned in tables \ref{tab1a}, \ref{tab1b}, \ref{tab2a} and \ref{tab2b} concern large (say 3 to 10mm) air bubbles in water for which the particulate Reynolds number is quite high ($\sim 800-2100$): they thus all correspond to bubbles in the same regime. The polydispersity is also significant in the experiments presented in tables tables \ref{tab1a}, \ref{tab1b}, \ref{tab2a} and \ref{tab2b}: the parameter $std(d)/d$ is not often quantified, but available measured size distributions indicate that this parameter keeps the same magnitude even though coalescence efficiency varies. In particular, let us underline that the flow conditions in tables \ref{tab1a}, \ref{tab1b}, \ref{tab2a} and \ref{tab2b} never concern bubbles whose size becomes of the order of the bubble column diameter (in other words, the flow conditions never correspond to slug flow). 

Thus, over the conditions mentioned in tables \ref{tab1a}, \ref{tab1b}, \ref{tab2a} and \ref{tab2b} that almost exclusively concern high aspect ratio bubble columns operated with a few millimeters in size air bubbles in water under ambient $T$, $P$ conditions, only two non-dimensional parameters have been significantly varied, namely:

\begin{itemize}
\item	the Archimedes number $Ar=gD^3\rho_L \delta \rho / \mu_L^2=[gD^3/\nu_L^2](\delta \rho/\rho_L)$
\item	the Froude number Froude number $Fr = V_{sg}/(gD)^{1/2}$.
\end{itemize}

Note that the Archimedes number equals $Re^2$, where $Re$ is the Reynolds number based on the velocity scale $(gD)^{1/2}$, on the size $D$ of the column and on the viscosity $\nu_L$ of the liquid. For the data shown Figure \ref{fig8}, $Ar$ ranges from $9.8  \times 10^9$ to $2.6 \times 10^{14}$ when in the heterogeneous regime. The Archimedes number $Ar$ is the equivalent of the Grashof number used in thermal convection where changes in density arise from differences in temperature and from fluid dilation instead of differences in local void fraction. In free thermal convection, the transition from laminar to turbulent regime corresponds to a Grashof number about $10^9$ (\cite{metais1964forced}). The magnitude of the Archimedes number in the heterogeneous regime discussed above exceeds indeed that critical Grashof limit: the heterogeneous regime in a bubble column can be seen as equivalent to the turbulent regime in thermal free convection. 

Owing to that observation, one could be tempted to associate the homogeneous/heterogeneous transition with a critical Grashof or Archimedes number. However, the above analysis was achieved in the asymptotic limit of a large aspect ratio. As far as the transition is concerned, both the column height (more precisely the static liquid height) and its diameter do affect the transition, as demonstrated by \cite{ruzicka2001effect}. Consequently, the parameter $Ho$ should be accounted for when discussing the transition, and the Grashof/Archimedes numbers definition should be adapted accordingly. Under such conditions, and as shown by \cite{ruzicka2003buoyancy}, the homogeneous/heterogeneous transition can be seen as an equivalent of thermal layers instability those transition is driven by a Rayleigh number. 

\section{\NOTE{Towards a 1D model to evaluate the void fraction}}\label{appB}

\begin{figure}
\centering
\includegraphics[width=\textwidth]{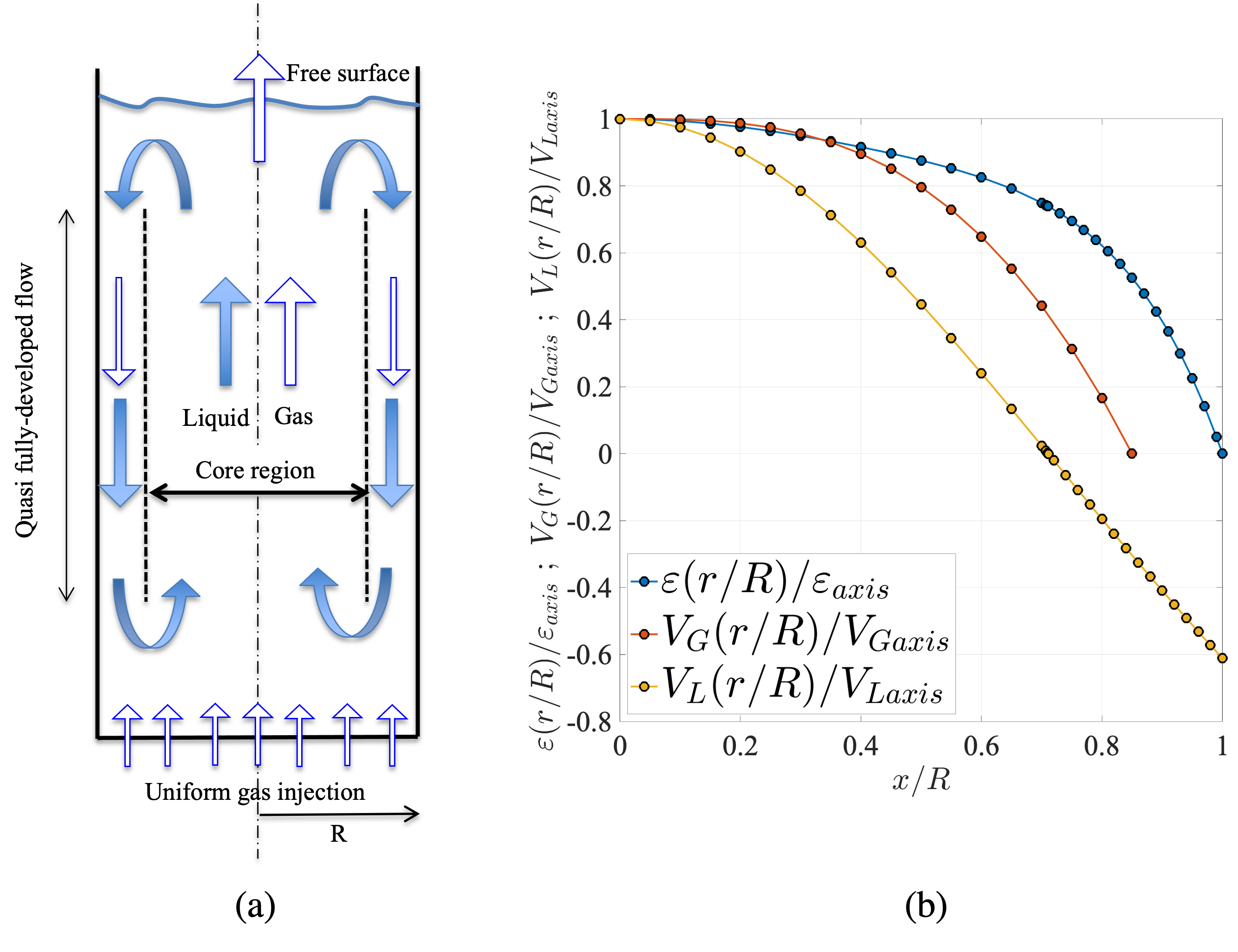}
\caption{\NOTE{(a): Sketch of the quasi one-dimensional two-phase flow in the core region of a bubble column operated in the heterogeneous regime. (b): transverses profiles of flow parameters normalized by their value on the axis versus $r/R$: the void fraction and the mean liquid velocity are the fits proposed by \cite{forret2006scale} and the mean gas velocity is the fit proposed by \cite{lefebvre2022new}. For the gas, the profile only concerns upward directed velocities, the downward directed gas velocity between $x/R=0.85$ and $x/R=1$ is not represented.}} \label{figB1}
\end{figure}

\NOTE{Our starting point is the same as the one exploited in \cite{raimundo2019hydrodynamics} to evaluate the apparent relative velocity between phases. In the core region of a bubble column operated in the heterogeneous regime, the mean liquid flow is directed upwards (Figure \ref{figB1}). Experiments indicate that this region has a radius comprised between 0.7R (\cite{kawase1986liquid}) and $0.71R$ (\cite{forret2006scale}). \cite{raimundo2019hydrodynamics} suggested that the actual limit may be the radius that equalizes the cross-section area of upflow and downflow regions i.e. ($2^{1/2}/2$) R. As we consider distances to injection such that the mean flow is steady and quasi fully developed (see the discussion in section  \ref{sec2}), let us assume that there is no radial exchange with the downward directed flow near walls so that the mean flow is purely one-dimensional.}

\NOTE{As the core zone forms a vertical inner tube of constant cross-section where the mean flow is one-dimensional, one can rely on kinematic models (\cite{zuber1965average}) to relate the void fraction to the gas flow rate fraction $\beta$. The latter is defined as $\beta= Q_{Gup}/(Q_{Gup} + Q_{Lup})$ where $Q_{Lup}$ (respectively $Q_{Gup}$) is the liquid (respectively gas) flow rate flowing upward in the tube. By definition, $Q_{Gup} = \langle \varepsilon U_G \rangle S_{core}$ and $Q_{Lup} = \langle(1 -\varepsilon) U_L \rangle S_{core}$, where $S_{core}$ is the cross section of the inner tube, and where the brackets $\langle . \rangle$ denote the spatial average over the cross section   (i.e. $\langle f \rangle$ equals the spatial integral of $f$ over the cross section divided by the area of that cross section). Following \cite{zuber1965average}, let us introduce the coefficients a and b that depend on the void fraction and on the phasic velocity profiles, namely: }

\begin{equation}\label{eqB1}
a = \langle \varepsilon U_G \rangle /  \left(  \langle \varepsilon  \rangle  \langle U_G \rangle  \right) ~;~ b= \langle (1- \varepsilon) U_L \rangle /  \left(  \langle 1- \varepsilon  \rangle  \langle U_L \rangle  \right) .
\end{equation} 
\\
\NOTE{Then, one has $Q_{Gup} = a \langle \varepsilon \rangle  \langle U_G \rangle S_{core}$ and $Q_{Lup} = b \langle 1-\varepsilon \rangle \langle U_L \rangle S_{core}$. Since by definition $Q_ {Gup}/Q_{Lup} = \beta /(1- \beta)$, one can write:}

\begin{equation}\label{eqB2}
\beta / (1-\beta) =  (a/b) \left( \langle U_G \rangle / \langle U_L \rangle \right) \left[ \langle \varepsilon \rangle / (1- \langle \varepsilon \rangle ) \right] = (a/b) \left( 1+ \langle U_R \rangle / \langle U_L \rangle \right) \left[ \langle \varepsilon \rangle / (1- \langle \varepsilon \rangle ) \right] .
\end{equation} 

\noindent \NOTE{That equation states that the difference between the gas concentration $\langle \varepsilon \rangle$ and the gas flow rate fraction $\beta$ is set by the ratio of the relative velocity of the gas averaged over the cross-section of the inner tube $\langle U_R \rangle$, to the liquid velocity $\langle U_L \rangle$ averaged over the same cross-section. As expected, a positive relative velocity leads to a concentration lower than the gas flow rate fraction.} 

\NOTE{From the known empirical profiles for void fraction and for mean liquid and gas velocities in the heterogeneous regime (see \ref{figB1}), one gets $a=1.02$ and $b=0.98$. These figures change by less than 1\% when the limit of the inner tube is varied from $0.7R$ to $0.71R$. As both coefficients a and b are very close to unity, and as they do not change with flow conditions, it is the ratio $\langle U_G \rangle/\langle U_L \rangle=1+\langle U_R \rangle/ \langle U_L \rangle$ that directly controls the proportionality between $\langle \varepsilon \rangle$ and $\langle \beta \rangle$. The ratio $\langle U_L \rangle /U_L(0)$ and $\langle U_G \rangle/U_G(0)$ evaluated from the known velocity profiles are given in Table \ref{tabB1}.}

\begin{table}
\begin{center}
\begin{tabular}{|c|c|c|c|c|}
\hline

  Inner tube radius: & $0.7R$ & $(2^{1/2}/2)R$ & $0.71R $ & Origin: \\
\hline
\hline
$\langle U_G \rangle / U_G(0)$ & 0.775 &  0.768  & 0.765  & fit proposed by \cite{lefebvre2022new} \\
\hline
$\langle U_L \rangle / U_L(0)$ & 0.472 & 0.463   & 0.459  & fit proposed by \cite{forret2006scale}  \\
\hline
\end{tabular}
\caption{\NOTE{Coefficients $\langle U_L \rangle /U_L(0)$ and $ \langle U_G \rangle/U_G(0)$ in the heterogeneous regime.}}\label{tabB1}
\end{center}
\end{table}

\NOTE{Hence, $(a/b) (\langle U_G \rangle/\langle U_L \rangle) = 1.73 U_G(0)/U_L(0)$ when the inner tube radius is set to $(2^{1/2}/2) R$, and eq. \ref{eqB2} can be rewritten:}

\begin{equation}\label{eqB3}
\beta / (1-\beta) =  1.73 U_G(0)/U_L(0) \left[ \langle \varepsilon \rangle / (1- \langle \varepsilon \rangle ) \right] .
\end{equation} 

\NOTE{\noindent This equation can be used to predict the void fraction if the gas flow rate fraction $\beta$ and $U_G(0)/U_L(0)$ are known. Alternately, one can estimate $\langle U_G \rangle / \langle U_L \rangle$ from the knowledge of $\beta$  and from void fraction measurements: this is how the relative velocity was evaluated in \cite{raimundo2019hydrodynamics}.}

\NOTE{To exploit eq. \ref{eqB2} or \ref{eqB3}, one needs to know the gas flow rate fraction $\beta$ relative to the inner tube, namely $\beta= Q_{Gup} /(Q_{Gup} + Q_{Lup})$ where both liquid $Q_{Lup}$ and gas $Q_{Gup}$ flow rates are evaluated within the inner tube.}

\begin{itemize}

\item	\NOTE{ For the liquid phase, the total upward liquid flow rate $Q_{Lup}$ is given by eq. \ref{eq6}, a result valid over a large range of flow conditions. }

\item	\NOTE{ For the gas phase, we argue that the gas flow rate flowing upward in the bubble column is equal to the flow rate injected over the entire column cross section, that is $Q_G = \pi R^2 V_{sg}$, plus some gas flow rate due to the global recirculation that re-entrain bubbles from the top to the bottom. We assume that the latter is a fraction $c$ of the injected gas flow rate, so that the total upward directed gas flow rate is $(1+c) Q_G$. }

\item	\NOTE{ Besides, the mean bubble velocity and void fraction profiles (Figure \ref{figB1}) indicate that the gas flows upwards between the axis and a radius equal to $\sim 0.85R$. That region is larger than the inner tube into which the liquid flows upward, and only the gas flux within the inner tube of radius $0.7$ to $0.71R$ must be counted when exploiting eq. \ref{eqB4}. According to known transverse profiles, the fraction of the gas flow rate flowing upward in the corona between $0.7$ or $0.71 R$ and $0.85 R$ represents between 8.2 and 9.2\% of the total upward gas flow rate. Hence, the actual gas flow rate $Q_{Gup}$ flowing through the inner tube is $91\% \pm 0.5\%$ of the total gas flow rate flowing upward. Therefore: }
\begin{equation}\label{eqB4}
Q_{Gup}=0.91(1+c)Q_G=0.91(1+c)\pi R^2 V_{sg}= 1.83  (1+c) / S_{core}V_{sg},
\end{equation} 
\NOTE{where the coefficient 1.83 is known within $\pm0.02$ and $S_{core}$ denotes the cross-section of a tube of radius $(2^{1/2}/2) R$. Therefore, for a bubble column operated the heterogeneous regime (with the mentioned restrictions on flow conditions), the actual gas flow rate fraction in the inner tube writes:} 
\begin{equation}\label{eqB5}
\beta=1.83 V_{sg} (1+c) / \left[ 1.83 V_{sg} (1+c) + Q_{Lup} /  S_{core} \right]= 1.83 Fr(1+c) / \left[ 1.83 Fr (1+c) + 0.098 \right].
\end{equation} 
\NOTE{Note that a direct quantification of c is not accessible because reliable bubble velocity measurements in near wall regions are lacking. Yet, according to the estimations made in \cite{lefebvre2022new}, $c$ should vary between 0 and 0.2. This result is expected to hold for all the flow conditions considered here (see sections \ref{sec3} and \ref{sec4}).}
\end{itemize}

\begin{figure}
\centering
\includegraphics[width=0.6\textwidth]{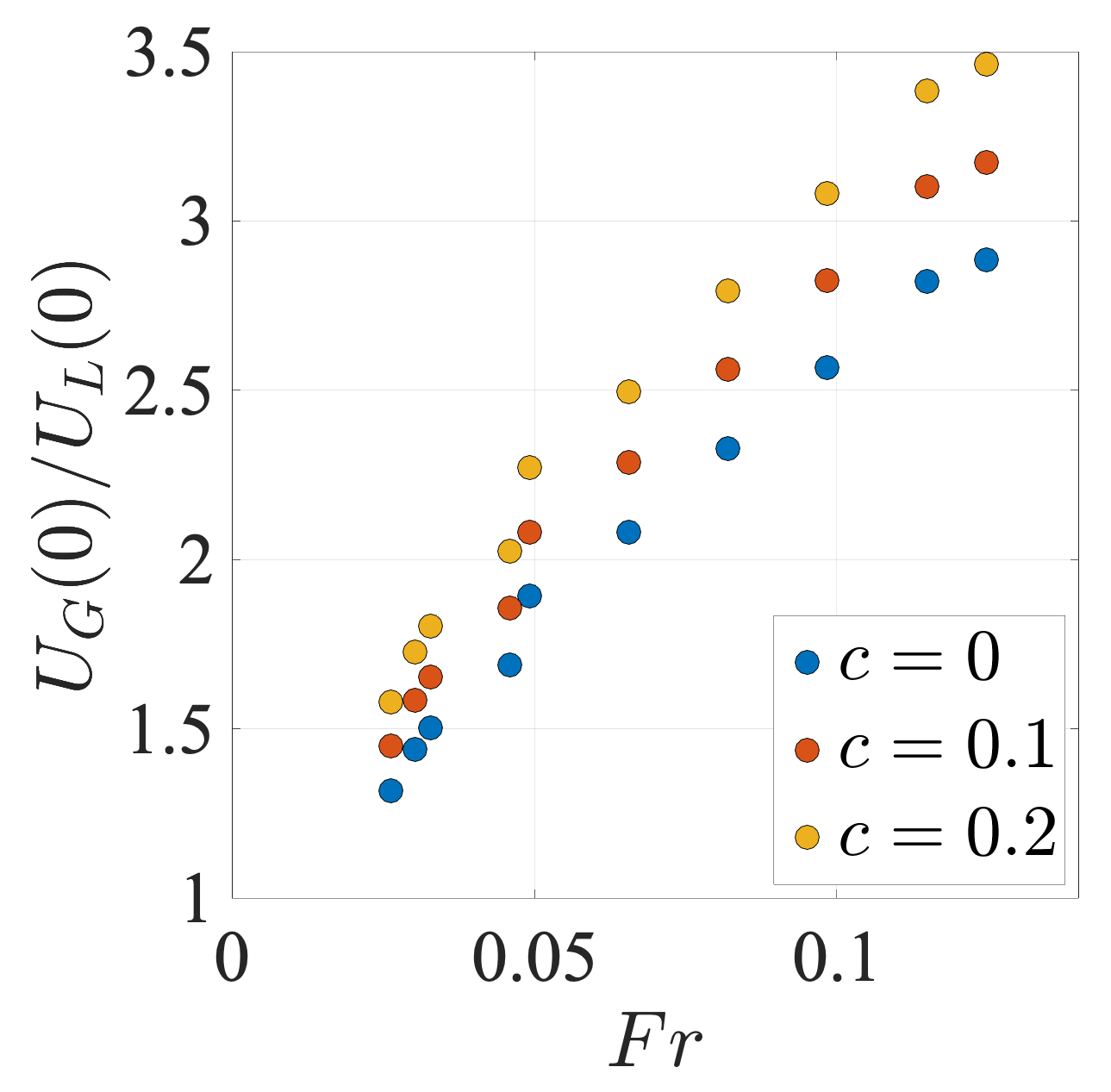}
\caption{\NOTE{Plot of the ratio $U_G(0)/U_L(0)$ deduced from eq. \ref{eqB3} and \ref{eqB5} versus $Fr$ for the data collected in the $D=0.4$m column at $H/D=3.625$ and in the heterogeneous regime.}} \label{figB2}
\end{figure}

\NOTE{Let us first reanalyse the data collected in the $D=0.4m$ column and in the heterogeneous regime. From the void fraction $\varepsilon_{axis}$ measured on the axis and for $H/D=3.625$ (Fig.\ref{fig3a}), the average void fraction $\varepsilon$ in the core region is estimated as $0.873 \varepsilon_{axis}$ according to the void fraction profile proposed by \cite{forret2006scale}. The ratio $U_G(0)/U_L(0)$ deduced from eq. \ref{eqB3} and eq. \ref{eqB5} is plotted versus $Fr$ in figure \ref{figB2}. Three values of $c$, namely 0 ; 0.1 and 0.2 have been considered: the results remain identical within 20\%. }

\NOTE{From the scalings given by eq. \ref{eq4} and \ref{eq5}, one expects $U_G(0)/U_L(0) \sim 1.65$. This value is indeed recovered in figure \ref{figB2} at the lowest Froude numbers shown, that is for superficial velocities at the beginning of the heterogeneous regime. }

\NOTE{A second observation is that $U_G(0)/U_L(0)$ is monotonously increasing with $V_{sg}$: it growths from 1.5 up to about 3 at the largest $Fr$ considered that is $Fr \sim 0.12$. This result contrasts with the direct measurements of the $V_L/(gD \varepsilon)^{1/2}$ and of $V_G/(gD \varepsilon)^{1/2}$ presented in Fig. \ref{fig6} that show that the ratio $V_G/V_L$ does not evolve with $V_{sg}$. A plausible reason for this apparent discrepancy is that $U_G(0)/U_L(0)$ is deduced from a 1D model that represents a strong idealization of the actual flow. In particular, the \cite{zuber1965average} approach exploited here requires the flow to be fully developed. However, we have shown (\cite{lefebvre2022new}) that in the quasi-fully developed region of a bubble column, the void fraction linearly increases with the height above the injector, and that this increase is larger than the hydrostatic contribution (the latter changes the volume of bubbles and thus the void fraction). Even more, the slope $d \varepsilon_{axis}/dH$ is zero in the homogeneous regime, while this slope linearly increases with $V_{sg}$ in the heterogeneous regime (see Fig.26 in  \cite{lefebvre2022new}). These observations indicate that the flow is evolving along a vertical, that is, even if self-similarity holds, the values of phasic velocities on the column axis are expected to change with height. In addition, and as shown by the vertical profiles of void fraction, these axial evolutions are significantly sensitive to the amount of gas injected. }

\begin{itemize}

\item	\NOTE{Consequently, an extra dependancy on $V_{sg}$, which is not present in the \cite{zuber1965average} approach by construction, is thus expected to intervene in the relationship between $\beta$ and $\langle \varepsilon \rangle$.}  
\item	\NOTE{Another consequence is that not enough attention has been paid so far to the axial evolutions of key variables, and new data are needed to characterise the axial changes. Such information will also be helpful to more precisely identify the experimental trends (see the discussion on dispersion related with Fig.\ref{fig30}).}

\end{itemize}

\begin{figure}
\centering
\includegraphics[width=0.6\textwidth]{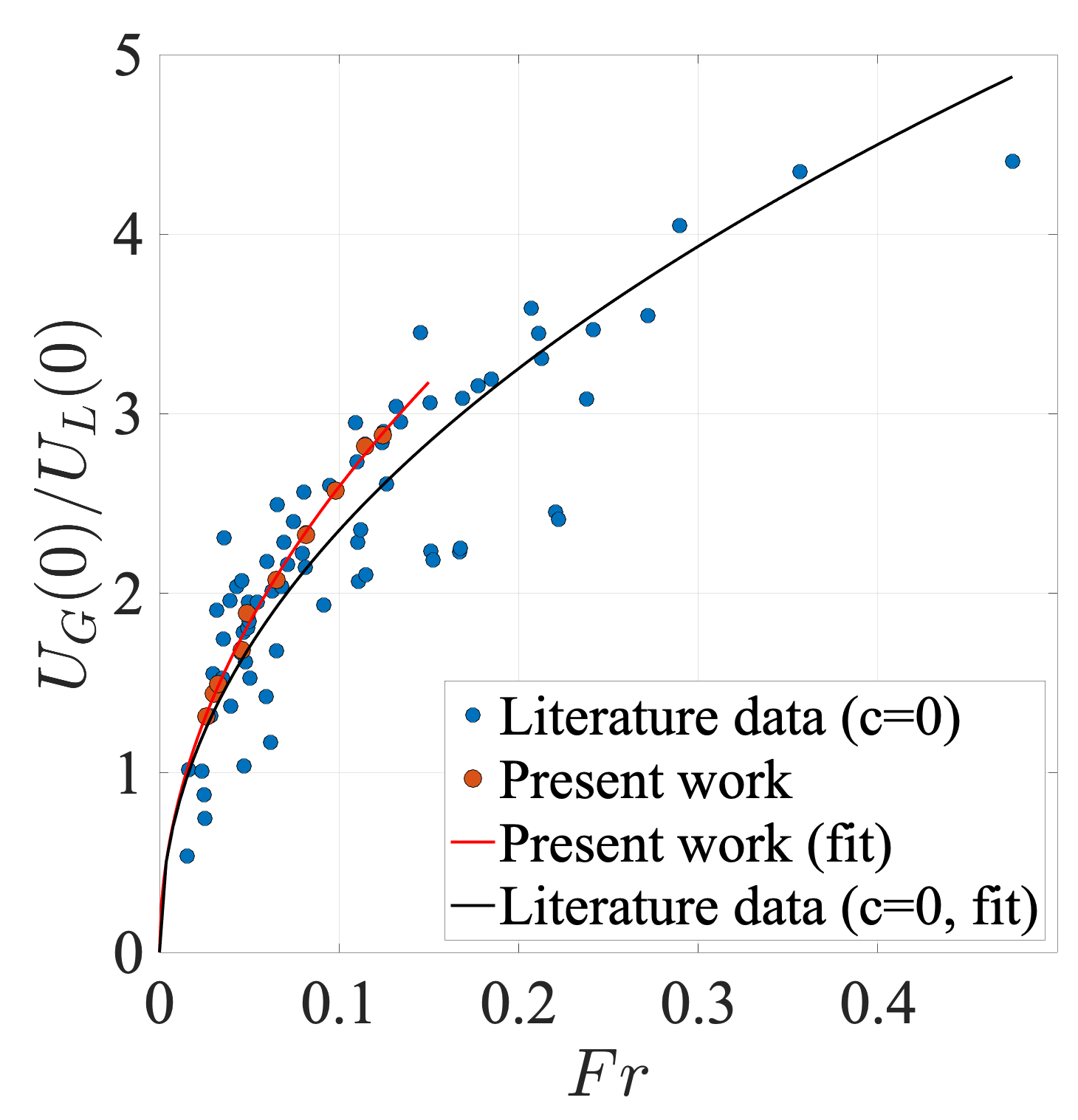}
\caption{\NOTE{ Plot of the ratio $U_G(0)/U_L(0)$ deduced from eq. \ref{eqB3} and \ref{eqB5} versus $Fr$ for datasets from literature mentioned in tables \ref{tab1a}, \ref{tab1b}, \ref{tab2a} and \ref{tab2b} in the heterogeneous regime (data within the transition region have been discarded and $c=0$ assumed). Data in red are those from the $D=0.4$m column. The fit (black solid line) for the datasets for the literature is $6.91 Fr^{0.47}$. The red solid line correspond to the fit to present's work data: $8.20 Fr^{0.50}$.}} \label{figB3}
\end{figure}

\NOTE{Similar conclusions arise from a reanalyse of data from literature. Fig. \ref{figB3} provides the ratio $U_G(0)/U_L(0)$ versus $Fr$ as deduced from heterogeneous regime data extracted from the articles quoted in tables \ref{tab1a}, \ref{tab1b}, \ref{tab2a} and \ref{tab2b} (that is for $ D $ between 0.138m and 3m, and for $V_{sg}$ from 4cm/s to 60cm/s). The trend is the same as the one identified for the $D=0.4$m column in Fig. \ref{figB2} . Let us also notice that the dispersion resulting from the selection of $c$ in the interval $[0 ; 0.2]$ is less than 20\%: this magnitude is comparable to the dispersion of void measurements observed when considering various heights above injection (see Fig.\ref{fig30}). The precise value of $c$ is therefore not critical.}

\NOTE{The increase of the ratio $U_G(0)/U_L(0)$ versus $Fr$ observed both in \ref{figB2} and in Fig. \ref{figB3} means that the pure 1D assumption needs to be relaxed. In particular, it is probable that the downward bubble flux along the walls is not entrained down to the bottom of the column in its entirety. Instead, some bubbles moving downward possibly feed the upflow motion all along the column height. That would continuously feed the upflow region at all altitudes with an extra bubble flux. If so, the gas flow rate fraction becomes a function of the altitude, and so does the void fraction $\varepsilon_{axis}$ as well as the transport velocities $V_L$ and $V_G$ on the axis. New experimental data are needed to test and characterize these options. }

\end{document}